\documentclass[journal]{IEEEtran}
\newif\ifpeerreview

\peerreviewfalse

\usepackage{xcolor}
\usepackage[switch]{lineno}
\usepackage{cite}
\usepackage{url}
\usepackage{lipsum}
\usepackage{graphicx}
\usepackage{float}
\usepackage{subfigure}
\usepackage{amsfonts}
\usepackage{amsmath}
\usepackage[ruled,noline,linesnumbered,noend]{algorithm2e}
\usepackage{booktabs}
\usepackage[bookmarks=false]{hyperref}

\usepackage{array}
\usepackage{graphbox}

\usepackage{float}

\usepackage{tikz}

\DeclareMathOperator*{\argmin}{arg\,min}

\begin{document}

\ifpeerreview
  \linenumbers
  \linenumbersep 5pt\relax
\fi

\title{Multi-Slice Fusion for Sparse-View and Limited-Angle 4D CT Reconstruction}
\ifpeerreview
\author{Anonymous submission}
\else
\author{
Soumendu~Majee,~\IEEEmembership{Student~Member,~IEEE},
Thilo~Balke,~\IEEEmembership{Student~Member,~IEEE},
Craig~A.~J.~Kemp,
Gregery~T.~Buzzard,~\IEEEmembership{Member,~IEEE}~and
Charles~A.~Bouman~\IEEEmembership{Fellow,~IEEE}

\thanks{Soumendu Majee, Thilo Balke, and Charles A. Bouman are with the School of Electrical and Computer Engineering, Purdue University, West Lafayette, IN, USA (e-mail: \href{mailto:smajee@purdue.edu}{smajee@purdue.edu}, \href{mailto:tbalke@purdue.edu}{tbalke@purdue.edu} , \href{mailto:bouman@purdue.edu}{bouman@purdue.edu}).
Craig A.J. Kemp is with Eli Lilly and Company, Indianapolis, IN, USA (email: \href{mailto:kemp_craig_a@lilly.com}{kemp\_craig\_a@lilly.com}).
Gregery T. Buzzard is with the Department of Mathematics, Purdue University, West Lafayette, IN, USA (email: \href{mailto:buzzard@purdue.edu}{buzzard@purdue.edu}).

This work was supported by Eli Lilly and Company under research project funding agreement
17099289.
Charles A. Bouman and Gregery T. Buzzard were supported in part by NSF grant CCF-1763896.
}
}
\fi

\IEEEtitleabstractindextext{%
\begin{abstract}

Inverse problems spanning four or more dimensions such as space, time and other independent parameters have become increasingly important.
State-of-the-art 4D reconstruction methods use model based iterative reconstruction (MBIR), but depend critically on the quality of the prior modeling.
Recently, plug-and-play (PnP) methods have been shown to be an effective way to incorporate advanced prior models using state-of-the-art denoising algorithms.
However, state-of-the-art denoisers such as BM4D and deep convolutional neural networks (CNNs) are primarily available for 2D or 3D images and extending them to higher dimensions is difficult due to algorithmic complexity and the
increased difficulty of effective training.

In this paper, we present {\em multi-slice fusion}, a novel algorithm for 4D reconstruction, based on the fusion of multiple low-dimensional denoisers. 
Our approach uses multi-agent consensus equilibrium (MACE), an extension of plug-and-play, as a framework for integrating the multiple lower-dimensional models.
We apply our method to 4D cone-beam X-ray CT reconstruction for non destructive evaluation (NDE) of samples that are dynamically moving during acquisition.
We implement multi-slice fusion on distributed, heterogeneous clusters in order to reconstruct large 4D volumes in reasonable time and demonstrate the inherent parallelizable nature of the algorithm.
We present simulated and real experimental results on sparse-view and limited-angle CT data to demonstrate that multi-slice fusion can substantially improve the quality of reconstructions relative to traditional methods, while also being practical to implement and train.

\end{abstract}

}

\maketitle
\IEEEdisplaynontitleabstractindextext

\section{Introduction}

Improvements in imaging sensors and computing power have made it possible to solve increasingly difficult reconstruction problems. 
In particular, the dimensionality of reconstruction problems has increased from the traditional~2D and 3D problems representing space
to more difficult 4D or even 5D problems representing space-time and, for example, heart or respiratory phase \cite{5D_huang2014mr,mohan2015timbir,balke2018separable,MSF_iccp,zeeshan,majee2017model}.

These higher-dimensional reconstruction problems pose surprisingly difficult challenges computationally and perhaps more importantly, in terms of algorithmic design and training due to the curse of dimensionality \cite{ziabariCNN}.
However, the high dimensionality of the reconstruction also presents important opportunities to improve reconstruction quality by exploiting the regularity in the high-dimensional space.
In particular, for time-resolved imaging, we can exploit the regularity of the image to reconstruct each frame with fewer measurements and thereby increase temporal resolution.
In the case of 4D CT, the contributions of
\cite{mohan2015timbir,gibbs2015three,zang2018space} have increased the temporal resolution by an order of magnitude by exploiting the space-time regularity of objects being imaged.
These approaches use model-based iterative reconstruction (MBIR) \cite{kisner2012model,sauer1993local} to enforce regularity in 4D using simple space-time prior models.
More recently, deep learning based post-processing for 4D reconstruction has been proposed as a method to improve reconstructed image quality~\cite{clark2019convolutional}.

\begin{figure}[!htb]
\centering     
\includegraphics[width=0.50\textwidth]{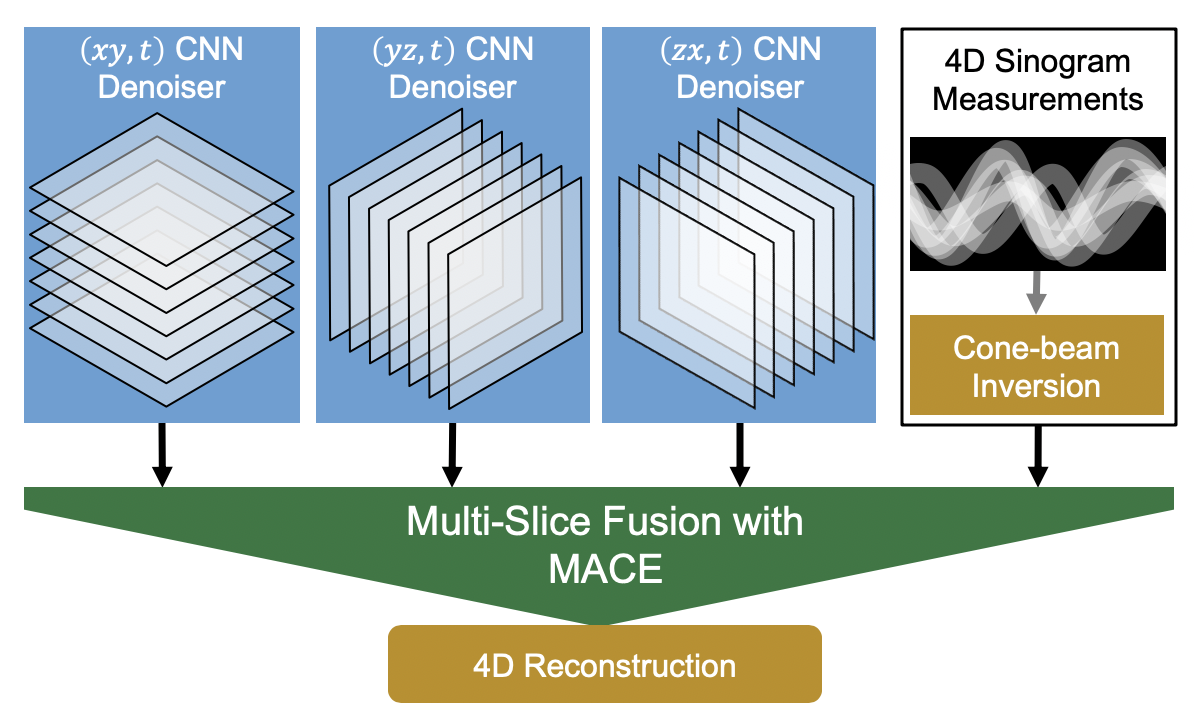}
\caption{Illustration of our multi-slice fusion approach.
Each CNN denoiser operates along the time direction and two spatial directions.
We fuse the CNN denoisers with the measurement model to produce a 4D regularized reconstruction.}
\label{fig:multislice_fusion}
\end{figure}

Recently, it has been demonstrated that plug-and-play (PnP) priors \cite{sreehari2016plug,venkatakrishnan2013plug,sun2018online,kamilov2017plug} can dramatically improve reconstruction quality by enabling the use of state-of-the-art denoisers as prior models in MBIR.
So PnP has great potential to improve reconstruction quality in 4D CT imaging problems. 
However, state-of-the-art denoisers such as deep convolutional neural networks (CNN) and BM4D are primarily available for 2D and sometimes 3D images,
and it is difficult to extend them to higher dimensions \cite{bm3d,bm4d,ziabariCNN}.
In particular, extending CNNs to 4D requires very computationally and memory intensive 4D convolution applied to 5D feature tensor structures.
This problem is further compounded by the lack of GPU accelerated routines for 4D convolution from major Deep-Learning frameworks such as Tensorflow, Keras, PyTorch~\footnote[1]{Currently only 1D, 2D, and 3D convolutions are supported with GPU acceleration}.
Furthermore, 4D CNNs require 4D ground truth data to train the PnP denoisers, which might be difficult or impossible to obtain.

In this paper, we present a novel 4D X-ray CT reconstruction algorithm that combines multiple low-dimensional CNN denoisers to implement a highly effective 4D prior model. 
Our approach, {\em multi-slice fusion}, integrates the multiple low-dimensional priors using multi-agent consensus equilibrium (MACE) \cite{buzzard2018plug}.
MACE is an extension of the PnP framework that formulates the inversion problem using an equilibrium equation---as opposed to an optimization---and allows for the use of multiple prior models and agents.

Figure~\ref{fig:multislice_fusion} illustrates the basic concept of our approach.
Multi-slice fusion integrates together three distinct CNN denoisers each of which is trained to remove additive white Gaussian noise along lower dimensional slices (hyperplanes) of the 4D object.
When MACE fuses the denoisers it {\em simultaneously} enforces the constraints of each denoising agent,
so that the reconstructions are constrained to be smooth in all four dimensions. 
Consequently, multi-slice fusion results in high-quality reconstructions that are practical to train and compute even when the dimensionality of the reconstruction is high.
In our implementation, one MACE agent estimates the cone-beam tomographic inversion.
The remaining 3 agents are CNN denoisers trained to remove additive white Gaussian noise along two spatial directions and the time direction.
The CNN agents work along complimentary spatial directions and are designed to take as input a stack of five 2D slices from five neighboring time-points.
We refer to this as 2.5D denoising~\cite{ziabariCNN,jiang2018denoising}.
Further details are given in Section~\ref{sec:MSF}.

The MACE solution can be computed using a variety of algorithms, including variants of the plug-and-play algorithm based on ADMM or other approaches \cite{sun2018plug,sun2018regularized,venkatakrishnan2013plug,sreehari2016plug}.
We implement multi-slice fusion on distributed heterogeneous clusters in which different agent updates are distributed onto different cluster nodes.
In particular, the cone-beam inversion computations are distributed onto multiple CPU nodes and concurrently, the CNN denoising computations are distributed onto multiple GPU nodes.

We present experiments using both simulated and real data of 4D NDE tomographic imaging from sparse-views, and we compare multi-slice fusion with MBIR using total variation (TV) and 4D Markov random field (MRF) priors. Our results indicate that multi-slice fusion can substantially reduce artifacts and increase resolution relative to these alternative reconstruction methods.

The rest of the paper is organized as follows.
In section~\ref{sec:problem}, we introduce the problem of 4D CT reconstruction.
In section~\ref{sec:MACE}, we introduce the theory behind MACE model fusion.
In section~\ref{sec:MSF}, we use the MACE framework to introduce multi-slice fusion.
In section~\ref{sec:train}, we describe our training pipeline for training the CNN denoisers.
In section~\ref{sec:parallel}, we describe our distributed implementation of multi-slice fusion on heterogeneous clusters.
Finally, in section~\ref{sec:results}, we present results on sparse-view and limited-angle 4D CT using both simulated and real data.

\section{Problem Formulation}\label{sec:problem}

\begin{figure}[!htb]
\centering     
\includegraphics[trim={0cm 0.5cm 0cm 0.5cm},clip,width=0.41\textwidth]{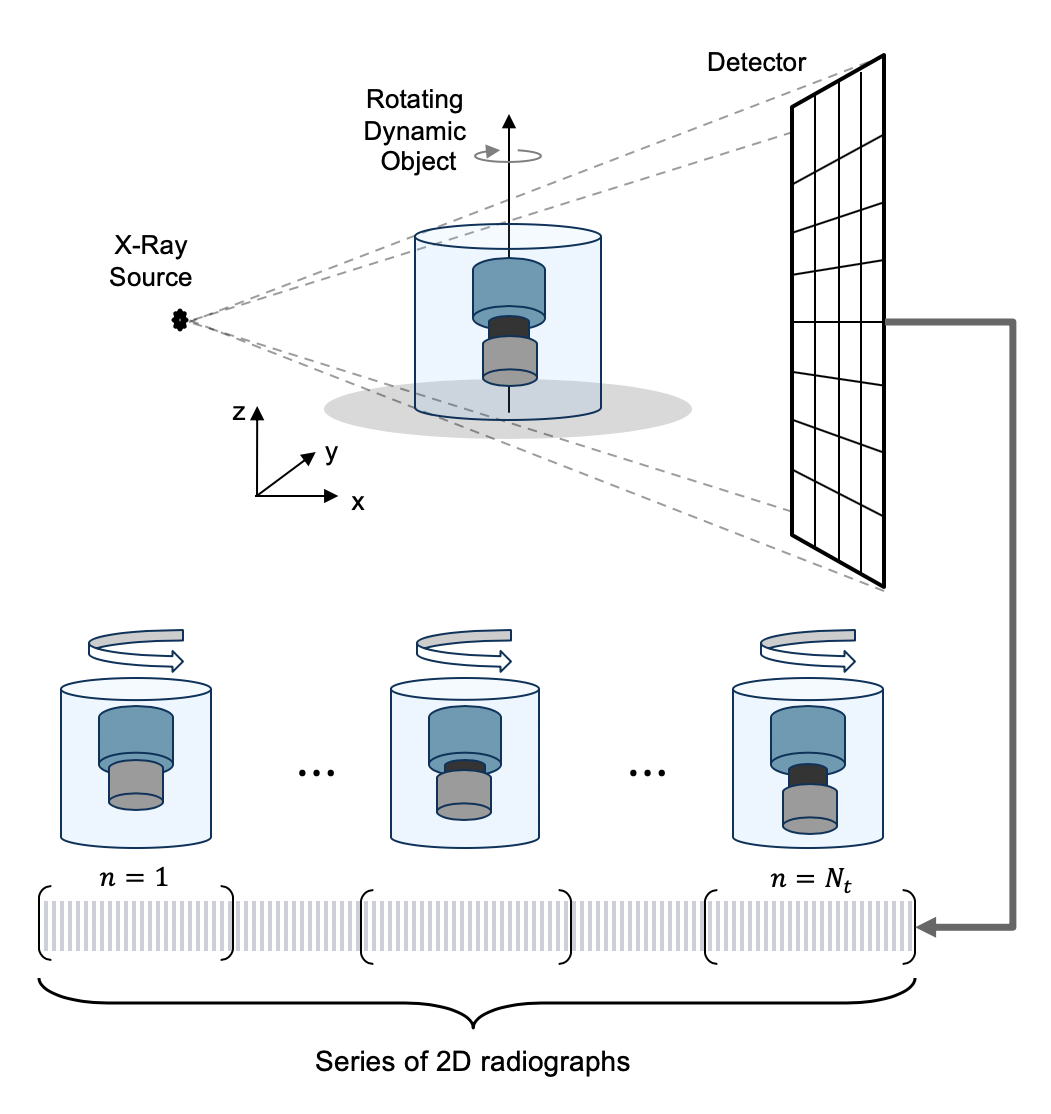}
\caption{Illustration of 4D cone-beam X-ray CT imaging.
The dynamic object is rotated and several 2D projections (radiographs) of the object are measured for different angles.
The projections are divided into $N_t$ disjoint subsets for each of the $N_t$ time-points.
}
\label{fig:conebeam}
\end{figure}

In 4D X-ray CT imaging, a dynamic object is rotated and several 2D projections (radiographs) of the object are measured for different angles as illustrated in Figure~\ref{fig:conebeam}.
The problem is then to reconstruct the 4D array of X-ray attenuation coefficients from these measurements, where three dimensions correspond to the spatial dimensions and the fourth dimension corresponds to time.

Let $N_t$ be the number of time-points, $M_n$ be the number of measurements at each time-point, and $N_s$ be the number of voxels at each time-point of the 4D volume.
For each time-point $n \in \{1,\hdots,N_t\}$, define $y_n \in \mathbb{R}^{M_n} $ to be the vector of sinogram measurements at time $n$, and $x_n \in \mathbb{R}^{N_s}$ to be the vectorized 3D volume of X-ray attenuation coefficients for that time-point.
Let us stack all the measurements to form a measurement vector $ y = [ y_1^\top, .. , y_{N_t}^\top ]^\top \in \mathbb{R}^{M} $ where $M = \sum_{n=1}^{N_t} M_n $ is the total number of measurements.
Similarly, let us stack the 3D volumes at each time-point to form a vectorized 4D volume $ x = [ x_1^\top, \hdots , x_{N_t}^\top  ]^\top \in \mathbb{R}^{N} $, where $N = N_t N_s$ is the total number voxels in the 4D volume.
The 4D reconstruction problem then becomes the task of recovering the 4D volume of attenuation coefficients, $x$, from the series of sinogram measurements, $y$.

In the traditional maximum a posteriori (MAP) approach, the reconstruction is given by
\begin{linenomath*}
\begin{equation}\label{eq:map_est}
    x^{*} = \argmin_{x} \left\{ l(x) + \beta h(x) \right\} \ ,
\end{equation}
\end{linenomath*}
where $l(x)$ is the data-fidelity or log-likelihood term, $h(x)$ is the 4D regularizer or prior model, and the unit-less parameter $\beta$ controls the level of regularization in the reconstruction.
The data-fidelity term, $l(x)$, can be written in a separable fashion as
\begin{linenomath*}
\begin{equation}\label{eq:LL_time_expand}
    l(x) = \frac{1}{2} \sum_{n=1}^{N_t} \| y_n - A_n x_n \|_{\Lambda_n}^2  \ ,
\end{equation}
\end{linenomath*}
where $A_n$ is the system matrix, and $\Lambda_n$ is the weight matrix for time-point $n$.
The weight matrix accounts for the non-uniform noise variance due to a Gaussian approximation~\cite{bouman1996unified} of the underlying Poisson noise. The weight matrix is computed as $\Lambda_n = \text{diag}\{ c \exp \left\{ - y_n \right\}$ where the scalar $c$ is empirically chosen~\cite{mohan2015timbir}.

If the prior model, $h(x)$, can be expressed analytically like a 4D Markov random field (MRF) as in \cite{mohan2015timbir,MSF_iccp}, then the expression in equation~(\ref{eq:map_est}) can be minimized iteratively to reconstruct the image.
However, in practice, it can be difficult to represent an advanced prior model in the form of a tractable cost function $h(x)$ that can be minimized.
Consequently, PnP algorithms have been created as a method for representing prior models as denoising operations\cite{sreehari2016plug,venkatakrishnan2013plug}.
More recently, PnP methods have been generalized to the multi-agent consensus equilibrium (MACE) framework as a way to integrate multiple models in a principled manner~\cite{buzzard2018plug,MSF_iccp,sridhar2018distributed}.

\section{MACE Model Fusion}\label{sec:MACE}

In this section, we use the multi-agent consensus equilibrium (MACE) framework to fuse the data-fidelity term and multiple denoisers; these multiple denoisers form a single prior model for reconstruction.
This allows us to construct a 4D prior model using low-dimensional CNN denoisers (described in  Section~\ref{sec:MSF}).

To introduce the concept of consensus equilibrium, let us first consider a variation of the optimization problem in equation~\eqref{eq:map_est} with $K$ regularizers $h_k(x)$, $k=1,\hdots,K$.
The modified optimization problem can thus be written as
\begin{equation}
\label{eq:map_multireg}
    x^{*} = \argmin_{x} \left\{ l(x) + \dfrac{\beta}{K} \sum_{k=1}^{K} h_k(x) \right\} \ ,
\end{equation}
where the normalization by $K$ is done to make the regularization strength independent of the number of regularizers.

Now we transform the optimization problem of equation~\eqref{eq:map_multireg} to an equivalent consensus equilibrium formulation.
However, in order to do this, we must introduce additional notation.
First, we define the proximal maps of each term in equation~\eqref{eq:map_multireg}.
We define $L(x) : \mathbb{R}^N \rightarrow \mathbb{R}^N $ to be the proximal map of $l(x)$ as
\begin{equation}\label{eq:mace_L}
    L(x) = \argmin_{z \in \mathbb{R}^{N}} \left\{ l(z) + \frac{1}{2 \sigma^2} \| x-z \|_2^2 \right\} \ ,
\end{equation}
for some $\sigma > 0$.
Similarly, we define $H_k(x) : \mathbb{R}^N \rightarrow \mathbb{R}^N $ to be the the proximal map of each $h_k(x)$ , $k=1,\hdots,K$ as
\begin{equation}\label{eq:mace_H}
    H_k(x) = \argmin_{z \in \mathbb{R}^{N}} \left\{ \frac{1}{2 \sigma^2} \| x-z \|_2^2 + h_k(z) \right\} \ .
\end{equation}
Each of these proximal maps serve as agents in the MACE framework.
We stack the agents together to form a stacked operator $F : \mathbb{R}^{(K+1)N} \rightarrow \mathbb{R}^{(K+1)N} $ as
\begin{equation}\label{eq:mace_F}
    F(W) = \left[ \begin{smallmatrix}  
    L(W_0 ) \\    
    H_1(W_1) \\ 
    \\
    \vdots
    \\
    \\
    H_K(W_K) \\
    \end{smallmatrix}\right] \ ,
\end{equation}
where $W \in \mathbb{R}^{(K+1)N} $ is stacked representative variable.
The consensus equilibrium is the vector $W^* \in \mathbb{R}^{(K+1)N}$ that satisfies
\begin{equation}\label{eq:mace_FG}
    F(W^*) = G(W^*) \ ,
\end{equation}
where $G$ is an averaging operator given as
\begin{equation}\label{eq:mace_G}
    G(W) = \left[ \begin{smallmatrix}
    \overline{W} \\ 
    \vdots \\ \\ 
    \overline{W} \end{smallmatrix}\right] \ ,
\end{equation}
and the weighted average is defined as 
\begin{equation}\label{eq:mace_avg}
    \overline{W} =  \frac{1}{1+\beta} W_0 + \frac{\beta}{1+\beta} \left( \frac{1}{K} \sum_{k=1}^{K} W_k  \right)  \ .
\end{equation}
Notice the weighting scheme is chosen to balance the forward and prior models.
The unitless parameter $\beta$ is used to tune the weights given to the prior model and thus the regularization of the reconstruction.
Equal weighing of the forward and prior models can be achieved using $\beta=1$.

If $W^*$ satisfies the consensus equilibrium condition of equation~\eqref{eq:mace_FG}, then it can be shown~\cite{buzzard2018plug} that $\overline{W^*}$ is the solution to the optimization problem in equation~\eqref{eq:map_multireg}.
Thus if the agents in MACE are true proximal maps then the 
consensus equilibrium solves an equivalent optimization problem.

\begin{figure}[t]
\centering     
\includegraphics[trim={0 0.8cm 0 0.0cm},clip,width=0.3\textwidth]{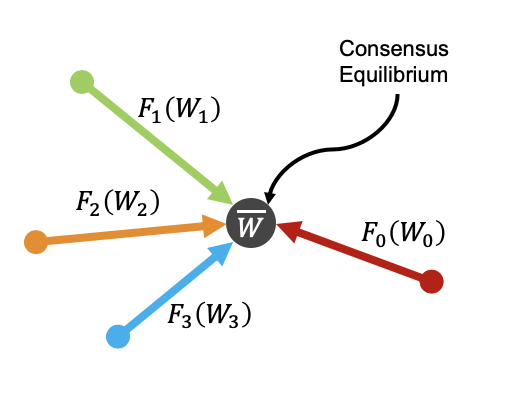}
\caption{Illustration of consensus equilibrium as analogous to a force balance equation: each agent pulls the solution toward its manifold and at equilibrium the forces balance each other.}
\label{fig:forceBalance}
\end{figure}

However, if the MACE agents are not true proximal maps, then there is no inherent optimization problem to be solved, but the MACE solution still exists.
In this case, the MACE solution can be interpreted as the balance point between the forces of each agent as illustrated in Figure~\ref{fig:forceBalance}.
Each agent pulls the solution toward its manifold  and the consensus equilibrium solution represents a balance point between the forces of each agent.
Thus MACE provides a way to incorporate non-optimization based models such as deep neural networks for solving inverse problems.

To see how we can incorporate deep neural network based prior models, first notice that equation~\eqref{eq:mace_H} can be interpreted as the MAP estimate for a Gaussian denoising problem with prior model $h_k$ and noise standard deviation $\sigma$.
Thus we can replace each MACE operator, $H_k$, for each $k=1,\hdots,K$ in equation~\eqref{eq:mace_H} with a deep neural network trained to remove additive white Gaussian noise of standard deviation $\sigma$.

It is interesting to note that when $H_k$ is implemented with a deep neural network denoiser,
then the agent $H_k$ is not, in general, a proximal map and there is no corresponding cost function $h_k$.
We know this because for $H_k$ to be a proximal map, it must satisfy the condition that $\nabla H_k(x) = [\nabla H_k(x)]^\top$ (see \cite{moreau1965proximite,sreehari2016plug}), which is equivalent to $H_k$ being a conservative vector function (see for example \cite[Theorem~2.6, p. 527]{vector_calculus_book}).
For a CNN, $\nabla H_k$ is a function of the trained weights, and in the general case, the condition will not be met unless the CNN architecture is specifically designed to enforce such a condition.


The consensus equilibrium equation~\ref{eq:mace_FG} states the condition that the equilibrium solution must satisfy.
However, the question remains of how to compute this equilibrium solution. 
Our approach to solving the consensus equilibrium equations is to first find an operator that has the equilibrium solution as a fixed point, and then use standard fixed point solvers. 
To do this, we first notice that the averaging operator has the property that $G(G(W)) = G(W)$. 
Intuitively, this is true because applying averaging twice is the same as applying it once. 
Using this fact, we see that
\begin{equation}
    (2G-I)(2G-I) = 4GG-4G+I = I \ ,
\end{equation}
where $I$ is the identity mapping.
We then rewrite equation~\eqref{eq:mace_FG} as 
\begin{eqnarray*}
F W^* &=& G W^* \\
(2F -I) W^* &=& (2G-I) W^* \\
(2G-I) (2F -I) W^* &=&  W^* \ .
\end{eqnarray*}
So from this we see that the following fixed point relationship must hold for the consensus equilibrium solution.
\begin{equation}\label{eq:mace_fixed_point}
    (2G-I) (2F -I) W^* =  W^* \ ,
\end{equation}
and the consensus equilibrium solution $W^*$ is a fixed point of the mapping $T = (2G-I)(2F-I)$.

We can apply a variety of iterative fixed point algorithms to equation~\eqref{eq:mace_fixed_point} to compute the equilibrium solution.
These algorithms have varying convergence guarantees and convergence speeds~\cite{buzzard2018plug}.
One such algorithm is Mann iteration~\cite{buzzard2018plug,sridhar2018distributed,sridharDistributed_CT_TCI}.
Mann iteration performs the following pseudo-code steps until convergence where $\leftarrow$ indicates assignment of a psuedo-code variable.
\begin{equation}\label{eq:mann}
    W \leftarrow  (1-\rho)W + \rho T W \ ,
\end{equation}
where weighing parameter $\rho \in (0,1)$ is used to control the speed of convergence.
In particular, when $\rho=0.5$, the Mann-iteration solver is equivalent to the consensus-ADMM algorithm~\cite{buzzard2018plug,boyd2011distributed}.
It can be shown that the Mann iteration converges to a fixed point of $T = (2G-I)(2F-I)$ if $T$ is a non-expansive mapping~\cite{buzzard2018plug}.

Note that each Mann iteration update in equation~\eqref{eq:mann} involves performing the minimization in equation~\eqref{eq:mace_L}.
This nested iteration is computationally expensive and leads to slow convergence.
Instead of minimizing equation~\eqref{eq:mace_L} till convergence, we initialize with the result of the previous Mann iteration and perform only three iterations of iterative coordinate descent (ICD).
We denote this partial update operator as $\tilde{L}(W_0 , X_0 )$ where $X_0$ is the initial condition to the iterative update. 
The corresponding new $F$ operator approximation is then given by
\begin{equation}\label{eq:L_tilde_operator}
   \tilde{F}(W;X) = \left[\begin{smallmatrix} 
   \Tilde{L}(W_{0};X_{0}) \\
    H_1(W_1) \\ 
    \\
    \vdots
    \\
    \\
    H_k(W_K) \\ 
    \end{smallmatrix}\right] \ .
\end{equation}

Algorithm~\ref{algo:CE_mann} shows a simplified Mann iteration using partial updates.
We perform algebraic manipulation of the traditional Mann iterations\cite{sridhar2018distributed,sridharDistributed_CT_TCI} in order to obtain the simplified but equivalent Algorithm~\ref{algo:CE_mann}.
It can be shown that partial update Mann iteration also converges~\cite{sridhar2018distributed,sridharDistributed_CT_TCI} to the fixed point in equation~\eqref{eq:mace_fixed_point}.
We used a zero initialization, $x^{(0)}=0$, in all our experiments and continue the partial update Mann iteration until the differences between state vectors $X_k$ become smaller than a fixed threshold.

\begin{algorithm}
\label{algo:CE_mann}
\DontPrintSemicolon
\KwIn{Initial Reconstruction: $ x^{(0)} \in \mathbb{R}^{N} $}
\KwOut{Final Reconstruction: $ x^{*} $}
$ X \leftarrow W \leftarrow
\left[\begin{smallmatrix} 
x^{(0)} \\
\vdots \\
x^{(0)}
\end{smallmatrix}\right] $ \;

\While{not converged}
{
    $ X \leftarrow \tilde{F}(W;X) $\;
    $ Z \leftarrow G( 2 X - W) $\;
    $ W \leftarrow W + 2 \rho(Z - X) $
}
$ x^{*} \leftarrow X_0 $

\caption{Partial update Mann iteration for computing the MACE solution
}
\end{algorithm}

\section{Multi-Slice Fusion using MACE}\label{sec:MSF}

\begin{figure}[t]
\centering     
\includegraphics[width=0.48\textwidth]{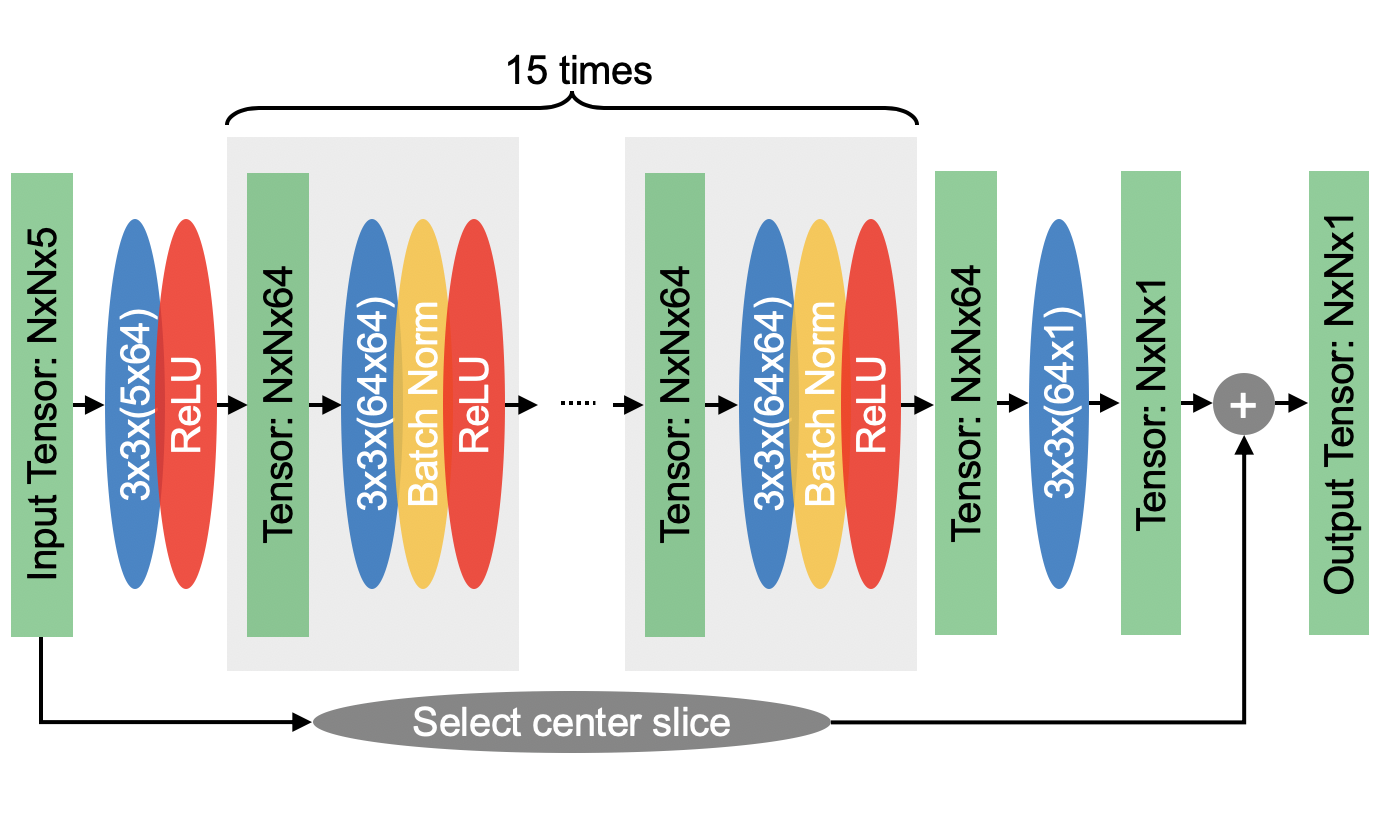}
\caption{Architecture of our 2.5D CNN denoiser.
Different sizes of input and output necessitate a selection operator for the residual connection.
Each green rectangle denotes a tensor, and each ellipse denotes an operation.
Blue ellipses specify the shape of the convolution kernel.
}
\label{fig:CNN}
\end{figure}

We use four MACE agents to implement multi-slice fusion.
We set $K=3$ and use the names $H_\text{xy,t}$, $H_\text{yz,t}$, $H_\text{zx,t}$ to denote the denoising agents $H_1$, $H_2$, $H_3$ in equation~\eqref{eq:mace_F}.
The agent $L$ enforces fidelity to the measurement while each of the denoisers $H_\text{xy,t}$, $H_\text{yz,t}$, $H_\text{zx,t}$ enforces regularity of the image in orthogonal image planes.
MACE imposes a consensus between the operators $L$, $H_\text{xy,t}$, $H_\text{yz,t}$, $H_\text{zx,t}$ to achieve a balanced reconstruction that lies at the intersection of the solution space of the measurement model and each of the prior models.
The MACE stacked operator $F$ encompassing all four agents can be written as
\begin{equation}\label{eq:MSF_F}
    F(W) = \left[\begin{smallmatrix}
    L(W_0) \\
    \\
    H_\text{xy,t}(W_1) \\ 
    \\
    H_\text{yz,t}(W_2) \\ 
    \\
    H_\text{zx,t}(W_3) \\ 
    \end{smallmatrix}\right] \ .
\end{equation}
Here the representative variable $W \in \mathbb{R}^{4N}$ is formed by stacking four vectorized 4D volumes.

The three denoisers $H_\text{xy,t}$, $H_\text{yz,t}$, and $H_\text{zx,t}$ share the same architecture and trained model but are applied along different planes of the 4D space.
The CNN architecture is shown in Figure~\ref{fig:CNN}.
We have modified a typical CNN architecture~\cite{dncnn} to input information from a third dimension.
The channel dimension of a convolution layer is typically used to input multiple color channels for denoising 2D color images using CNNs.
We re-purpose the channel dimension to input five adjacent 2D slices of the noisy image to the network and output the denoised center slice.
The other slices are being denoised by shifting the 5-slice moving window.
We call this 2.5D since the receptive field along the convolution dimensions is large but in the channel dimension is small.
It has been shown that this type of 2.5D processing is a computationally efficient way of performing effective 3D denoising with CNNs~\cite{ziabariCNN,jiang2018denoising}.
We use the notation $H_\text{xy,t}$ to denote a CNN space-time denoiser that performs convolution in the xy-plane and uses the convolution channels to input slices from neighboring time-points.
The denoisers $H_\text{yz,t}$ and $H_\text{zx,t}$ are analogous to $H_\text{xy,t}$ but are applied along the yz and zx-plane, respectively.
This orientation of the three denoisers ensures that
\begin{enumerate}
    \item The spatial dimensions x, y, z are treated equivalently. This ensures the regularization to be uniform across all spatial dimensions;
    \item Each dimension in x, y, z, and t is considered at least once. This ensures that model fusion using MACE incorporates information along all four dimensions.
\end{enumerate}

Since the three denoising operators $H_\text{xy,t}$, $H_\text{yz,t}$, and $H_\text{zx,t}$ process the 4D volume ``slice by slice'', they can be implemented in parallel on large scale parallel computers.
Details on distributed implementation are described in section~\ref{sec:parallel}.

\section{Training of CNN Denoisers} \label{sec:train}

All three prior model agents $H_\text{xy,t}$, $H_\text{yz,t}$, and $H_\text{zx,t}$ in multi-slice fusion share the same 2.5D model shown in Figure~\ref{fig:CNN} but are oriented along different planes.
Consequently we train a single 2.5D CNN model using 3D data.
Even though the CNN needs to denoise 3D time-space data, we train it using 3D spatial data since 3D volumes are widely available unlike time-space data.

\begin{figure}[!htb]
\centering     
\includegraphics[trim={0cm 0cm 0cm 0cm},clip,width=0.5\textwidth]{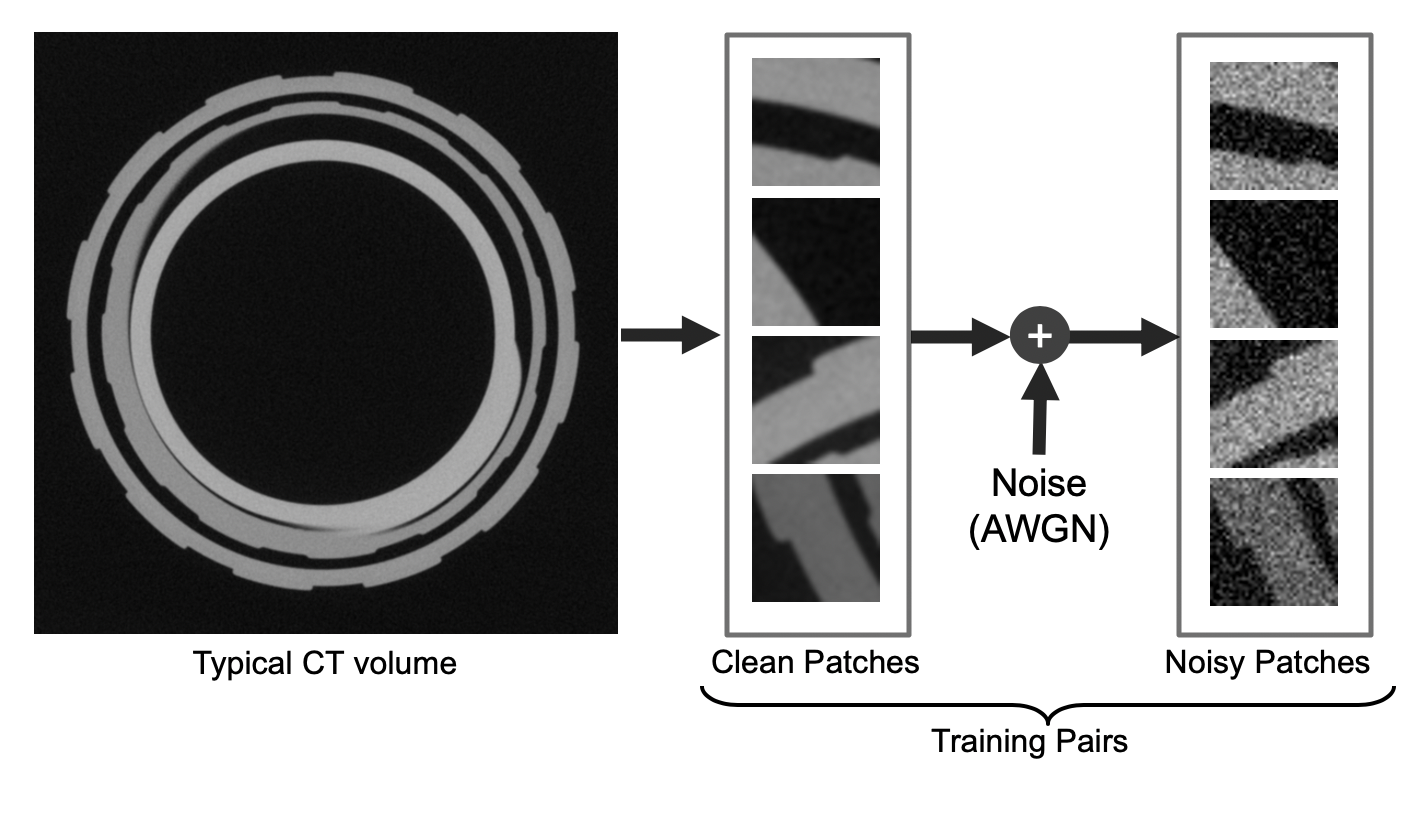}
\caption{Illustration of our training data generation.
We extract 3D patches from a typical CT volume and add additive white Gaussian noise (AWGN) to generate training pairs.
This makes the training process self-supervised.}
\label{fig:train}
\end{figure}

Figure~\ref{fig:train} outlines our training data generation.
We start with a low-noise 3D CT volume that is representative of the objects to be reconstructed.
We extract 3D patches from the CT volume and add pseudo-random additive white Gaussian noise (AWGN) to the patches to generate the training pairs.
We then train the CNN to remove the noise.
The use of AWGN is due to the mathematical form of the quadratic norm term in the proximal map in equation~\ref{eq:mace_H} and follows from the theory of Plug-and-play~\cite{sreehari2016plug,venkatakrishnan2013plug}.

\section{Distributed Reconstruction}\label{sec:parallel}

\begin{figure}[!ht]
\centering     
\includegraphics[width=0.48\textwidth]{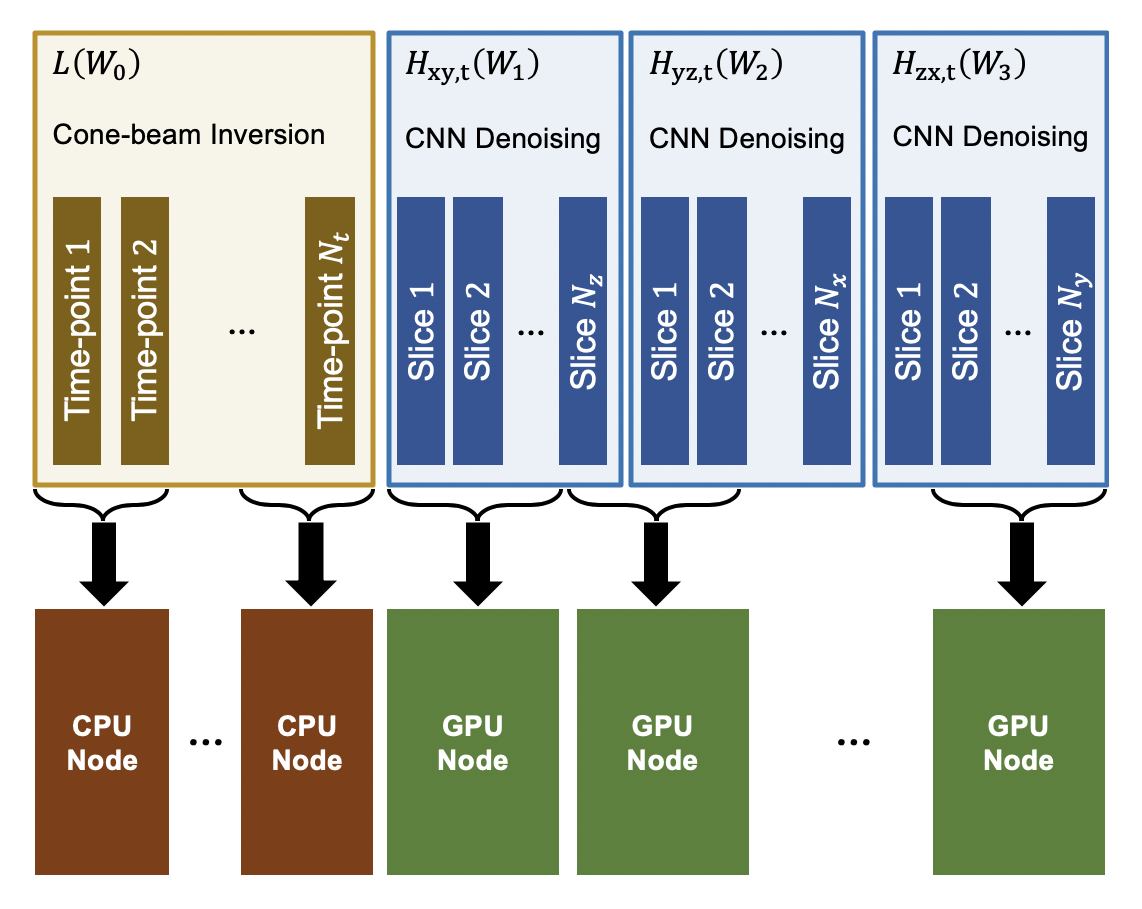}
\caption{Illustration of distributed computation of multi-slice fusion.
We perform distributed computation of the $F$ operator which is the main computational bottleneck in Algorithm~\ref{algo:CE_mann}.
Each operator within $F$, namely $H_\text{xy,t}$, $H_\text{yz,t}$, $H_\text{zx,t}$, and $L$ can be executed in parallel.
Furthermore, operators $H_\text{xy,t}$, $H_\text{yz,t}$, $H_\text{zx,t}$, and $L$ are 3D operators that can process the 4D volume ``slice by slice'' leading to a large number of concurrent operations that can be distributed among multiple compute nodes.
}
\label{fig:parallel}
\end{figure}

The computational structure of multi-slice fusion is well-suited to a highly distributed implementation.
The main computational bottleneck in Algorithm~\ref{algo:CE_mann} is the $F$ operator.
Fortunately, $F$ is a parallel operator and thus its individual components $L$, $H_\text{xy,t}$, $H_\text{yz,t}$, and $H_\text{zx,t}$ can be executed in parallel.
The operators $L$, $H_\text{xy,t}$, $H_\text{yz,t}$, and $H_\text{zx,t}$ can themselves be parallelized internally as well.
The distributed implementation of multi-slice fusion is illustrated in Figure~\ref{fig:parallel}.

The CNN denoisers $H_\text{xy,t}$, $H_\text{yz,t}$, and $H_\text{zx,t}$ are 2.5D denoisers that denoise the 4D volume by processing it slice by slice and thus can be trivially parallelized leading to a large number of concurrent operations.
The concurrent operations for all three denoisers are distributed among multiple GPUs due to the availability of optimized GPU routines in Tensorflow.
In our experiments we used a GPU cluster with three Nvidia Tesla P100 GPUs to compute the CNN denoising operators.

The cone-beam inversion operator, $L$, can also be computed for each time-point independently due to the separable structure in equations \eqref{eq:mace_L} and \eqref{eq:LL_time_expand}.
This leads to a large number of concurrent operations which are distributed among multiple CPU nodes.
The cone-beam inversion for each time-point is computed using a coordinate-descent minimization with multi-threaded parallelism.
Further details about the cone-beam inversion can be found in \cite{balke2018separable}.

\section{Experimental Results}\label{sec:results}

\begin{figure*}[!htb]
\centering     
\subfigure[Phantom]{\includegraphics[trim={1.4cm 1.8cm 5.8cm 1.8cm},clip,width=0.115\textwidth]
{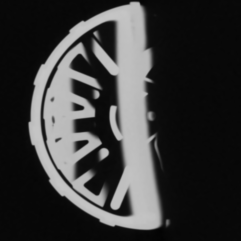}}
\subfigure[FBP (3D) ]{\includegraphics[trim={1.4cm 1.8cm 5.8cm 1.8cm},clip,width=0.115\textwidth]
{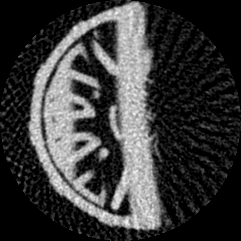}}
\subfigure[MBIR+TV (3D) ]{\includegraphics[trim={1.4cm 1.8cm 5.8cm 1.8cm},clip,width=0.115\textwidth]
{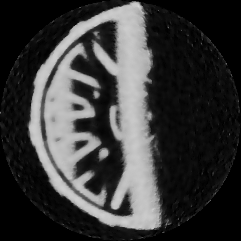}}
\subfigure[MBIR+4D-MRF ]{\includegraphics[trim={1.4cm 1.8cm 5.8cm 1.8cm},clip,width=0.115\textwidth]
{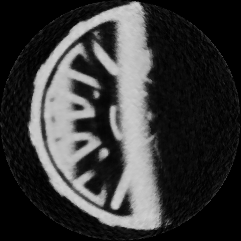}}
\subfigure[\textbf{Multi-slice fusion (proposed)}]
{
\begin{tikzpicture}
    \node[anchor=south west,inner sep=0] (image) at (0,0) {\includegraphics[trim={1.4cm 1.8cm 5.8cm 1.8cm},clip,width=0.115\textwidth]{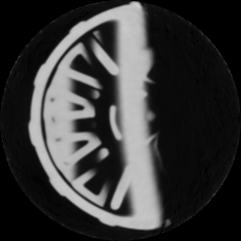}};
    \begin{scope}[x={(image.south east)},y={(image.north west)}]
        \draw[blue,ultra thick] (0.73,0.2) ellipse (0.15 and 0.035);
    \end{scope}
\end{tikzpicture}
}
\subfigure[MBIR+$H_\text{xy,t}$]{\includegraphics[trim={1.4cm 1.8cm 5.8cm 1.8cm},clip,width=0.115\textwidth]
{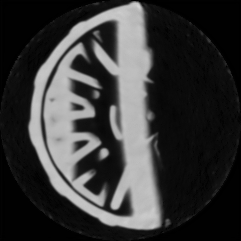}}
\subfigure[MBIR+$H_\text{yz,t}$]{\includegraphics[trim={1.4cm 1.8cm 5.8cm 1.8cm},clip,width=0.115\textwidth]
{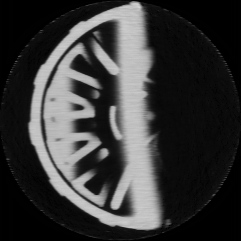}}
\subfigure[MBIR+$H_\text{zx,t}$]{\includegraphics[trim={1.4cm 1.8cm 5.8cm 1.8cm},clip,width=0.115\textwidth]
{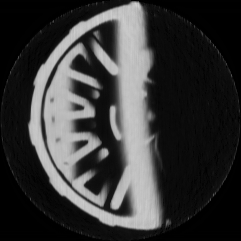}}
\caption{ Comparison of different methods for simulated data $360^\circ$.
Each image is a slice through the reconstructed object for one time-point along the spatial xy-plane.
The reconstruction using FBP suffers from high noise and fails to recover the small hole in the bottom of the image.
MBIR+TV and MBIR+4D-MRF suffer from jagged edges and fail to recover the small hole in the bottom of the image.
MBIR+$H_\text{yz,t}$ and MBIR+$H_\text{zx,t}$ suffer from horizontal and vertical streaks, respectively, since the denoisers were applied in those planes.
MBIR+$H_\text{xy,t}$ cannot reconstruct the small hole in the bottom of the image since the xy-plane does not contain sufficient information.
}
\label{fig:simresults_360_xy}
\end{figure*}

We present experimental results on two simulated and two real 4D X-ray CT data for Non-Destructive Evaluation (NDE) applications to demonstrate the improved reconstruction quality of our method.
The four experimental cases are outlined below
\begin{enumerate}
    \item \textbf{Simulated Data $\mathbf{360^\circ}$:} Sparse-view results on simulated data with a set of sparse views ranging over $360^\circ$ at each reconstructed time-point;

    \item \textbf{Simulated Data $\mathbf{90^\circ}$:} Sparse-view limited-angle results on simulated data with a set of sparse views ranging over $90^\circ$ at each reconstructed time-point;
    
    \item \textbf{Real Data $\mathbf{360^\circ}$:} Sparse-view results on real data with a set of sparse views ranging over $360^\circ$ at each reconstructed time-point;
    
    \item \textbf{Real Data $\mathbf{90^\circ}$:} Sparse-view limited-angle results on real data with a set of sparse views ranging over $90^\circ$ at each reconstructed time-point.
\end{enumerate}
The selection of the rotation range per time-point is arbitrary and can be chosen after the measurements have been taken. 
For example, a full rotation with 400 views can be used as a single time-point or as four time-points with 100 views each. 
The four time-points per rotation can provide extra temporal resolution, however, they require a more difficult reconstruction with incomplete information.

We compare multi-slice fusion with several other methods outlined below
\begin{itemize}
    \item FBP: Conventional 3D filtered back projection reconstruction;
    \item MBIR+TV: MBIR reconstruction using a total variation (TV) prior~\cite{getreuer2012rudin} in the spatial dimensions;
    \item MBIR+4D-MRF: MBIR reconstruction using 4D Markov random field prior~\cite{mohan2015timbir} with $q=2.2$, $p=1.1$, $26$ spatial neighbors and $2$ temporal neighbors;
    \item MBIR+$H_\text{xy,t}$: MBIR using the CNN $H_\text{xy,t}$ as a PnP prior;
    \item MBIR+$H_\text{yz,t}$: MBIR using the CNN $H_\text{yz,t}$ as a PnP prior;
    \item MBIR+$H_\text{zx,t}$: MBIR using the CNN $H_\text{zx,t}$ as a PnP prior.
\end{itemize}

We used two CPU cluster nodes, each with 20 Kaby Lake CPU cores and 96 GB system memory to compute the cone-beam inversion.
We used three GPU nodes, each with a Nvidia Tesla P100 GPU (16 GB GPU-memory) and 192 GB system memory to compute the CNN denoisers.
To compute the multi-slice fusion reconstruction, we run Algorithm~\ref{algo:CE_mann} for $10$ Mann iterations, with $3$ iterations of cone-beam inversion per Mann iteration.
The total reconstruction time of multi-slice fusion for each experimental case are given in Table~\ref{table:runtime}.

\begin{table}[h!]
\centering
\begin{tabular}{|l | c | r|} 
\hline
&&\\
Experimental Case & Reconstruction size (x,y,z,t)  & Reconstruction Time \\ 
&&\\
\hline
Simulated Data $360^\circ$ & $240 \times 240 \times 28 \times 8 $ & $8 \ \mathrm{mins}$  \\ 
Simulated Data $90^\circ$ & $240 \times 240 \times 28 \times 8 $ & $8 \ \mathrm{mins}$  \\
Real Data $360^\circ$ & $731 \times 731 \times 91 \times 16 $ & $133 \ \mathrm{mins}$   \\
Real Data $90^\circ$ & $263 \times 263 \times 778 \times 12 $ & $46 \ \mathrm{mins}$  \\
\hline
\end{tabular}
\\
\vspace{1mm}
\caption{Total reconstruction time of multi-slice fusion for each experimental case}
\label{table:runtime}
\end{table}

The 2.5D CNN denoiser model used in the reconstructions was trained using a low-noise 3D CT reconstruction of a bottle and screw cap made from different plastics.
The object is representative of a variety of Non-Destructive Evaluation (NDE) problems in which the objects to be imaged are constructed from a relatively small number of distinct materials.
The extracted patches were normalized to $[0,1]$ and random rotation, mirroring, intensity shift were applied.
The standard deviation of the additive white Gaussian noise added during training was $0.1$.

\begin{figure}[!htb]
\centering     
\subfigure[Phantom]{\includegraphics[width=0.08\textwidth]{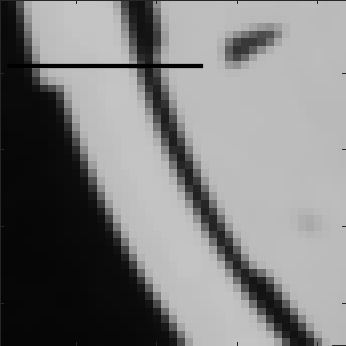}}
\subfigure[FBP (3D)]{\includegraphics[width=0.08\textwidth]{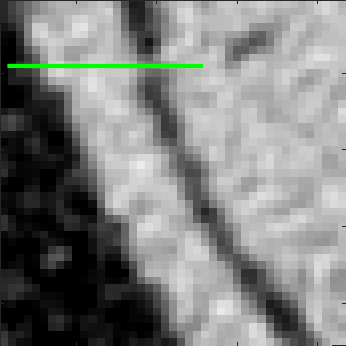}}
\subfigure[MBIR+TV (3D)]{\includegraphics[width=0.08\textwidth]{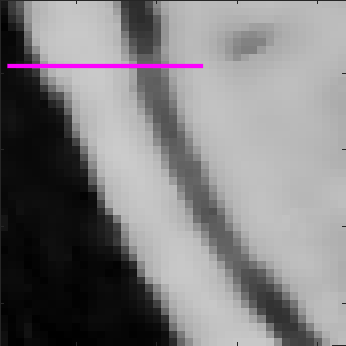}}
\subfigure[MIBIR+4D-MRF]{\includegraphics[width=0.08\textwidth]{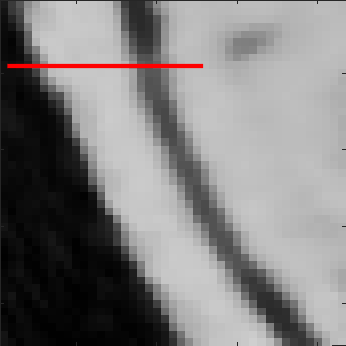}}
\subfigure[\textbf{Multi-slice fusion \newline (proposed)}]{\includegraphics[width=0.08\textwidth]{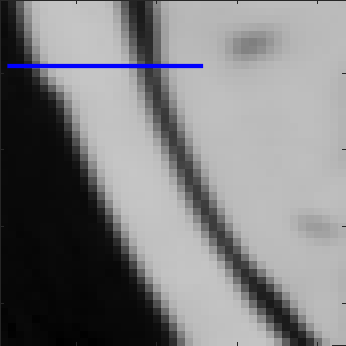}}
\subfigure[Plot of cross-section]{\includegraphics[width=0.5\textwidth]{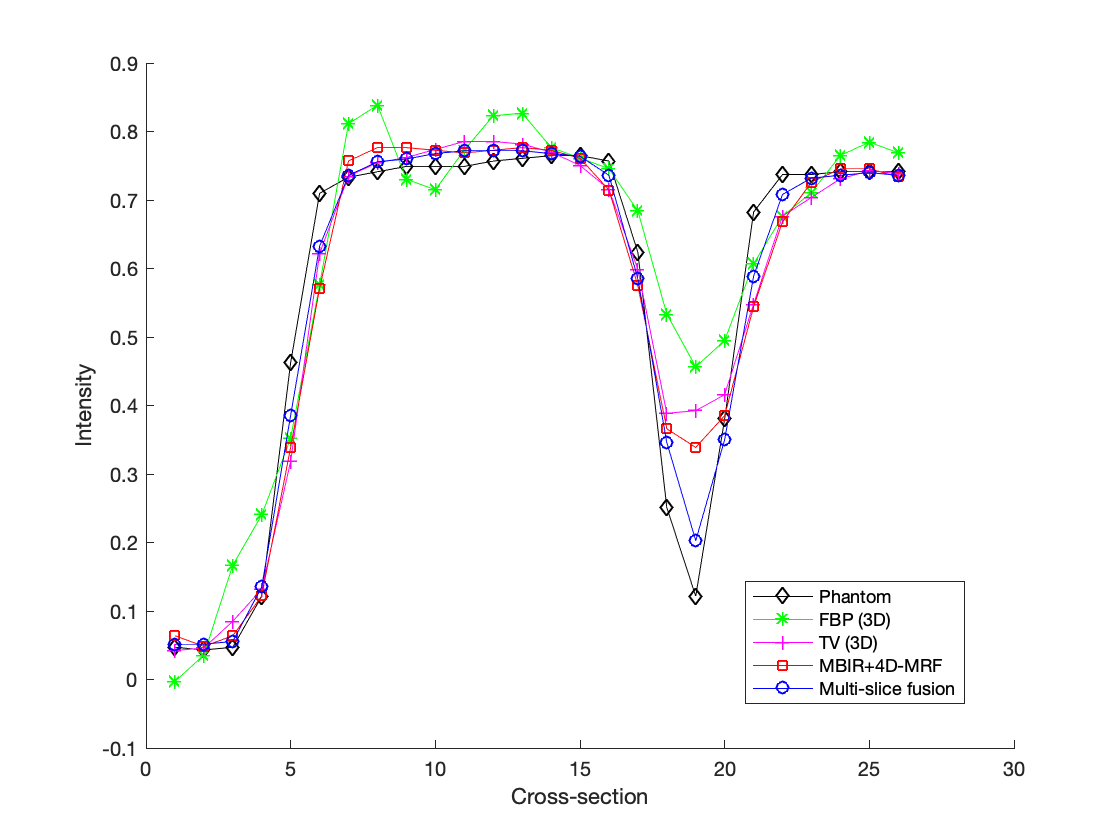}}
\caption{ Plot of cross-section through the phantom and reconstructions from simulated data $360^\circ$.
Multi-slice fusion results in the most accurate reconstruction of the gap between materials.}
\label{fig:simresults_360_crossSection}
\end{figure}

\begin{figure*}[!htb]
\centering     

\begin{center}
\begingroup
\setlength{\tabcolsep}{1.2pt} 
\renewcommand{\arraystretch}{0.2} 
\begin{tabular}{cccccc}
& &
Time-point 1 & Time-point 2 & Time-point 3 & Time-point 4\\
\\
\rotatebox[origin=c]{90}{Phantom} & &
\includegraphics[align=c,trim={0 0.5cm 2cm 0.5cm},clip,width=0.2\textwidth]{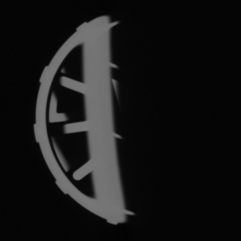}&
\includegraphics[align=c,trim={0 0.5cm 2cm 0.5cm},clip,width=0.2\textwidth]{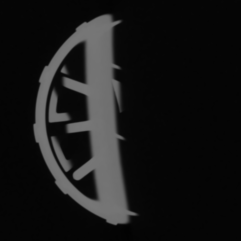}&
\includegraphics[align=c,trim={0 0.5cm 2cm 0.5cm},clip,width=0.2\textwidth]{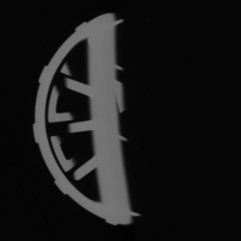}&
\includegraphics[align=c,trim={0 0.5cm 2cm 0.5cm},clip,width=0.2\textwidth]{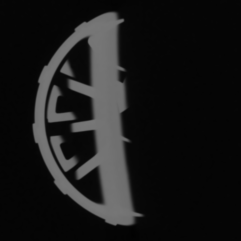}
\\
\\
\rotatebox[origin=c]{90}{FBP (3D)} & &
\includegraphics[align=c,trim={0 0.5cm 2cm 0.5cm},clip,width=0.2\textwidth]{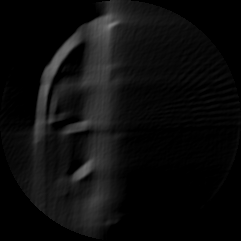}&
\includegraphics[align=c,trim={0 0.5cm 2cm 0.5cm},clip,width=0.2\textwidth]{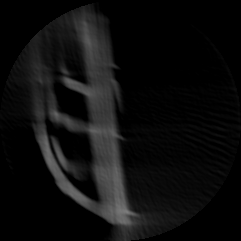}&
\includegraphics[align=c,trim={0 0.5cm 2cm 0.5cm},clip,width=0.2\textwidth]{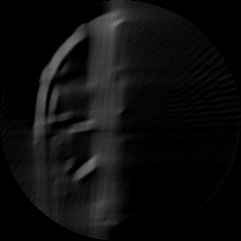}&
\includegraphics[align=c,trim={0 0.5cm 2cm 0.5cm},clip,width=0.2\textwidth]{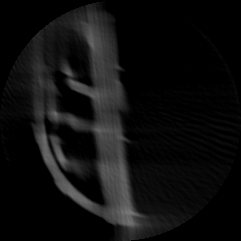}
\\
\\
\rotatebox[origin=c]{90}{MBIR+TV (3D)} & &
\includegraphics[align=c,trim={0 0.5cm 2cm 0.5cm},clip,width=0.2\textwidth]{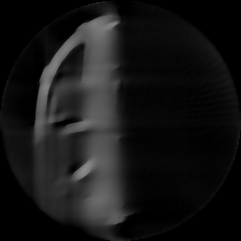}&
\includegraphics[align=c,trim={0 0.5cm 2cm 0.5cm},clip,width=0.2\textwidth]{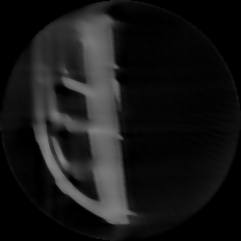}&
\includegraphics[align=c,trim={0 0.5cm 2cm 0.5cm},clip,width=0.2\textwidth]{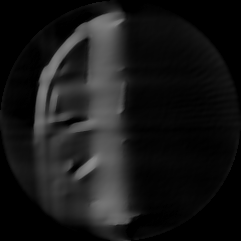}&
\includegraphics[align=c,trim={0 0.5cm 2cm 0.5cm},clip,width=0.2\textwidth]{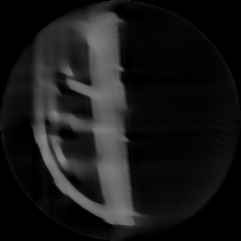}
\\
\\
\rotatebox[origin=c]{90}{MBIR+4D-MRF} & &
\includegraphics[align=c,trim={0 0.5cm 2cm 0.5cm},clip,width=0.2\textwidth]{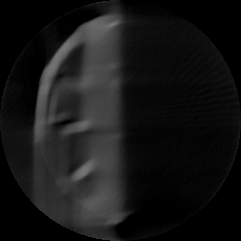}&
\includegraphics[align=c,trim={0 0.5cm 2cm 0.5cm},clip,width=0.2\textwidth]{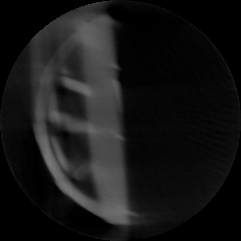}&
\includegraphics[align=c,trim={0 0.5cm 2cm 0.5cm},clip,width=0.2\textwidth]{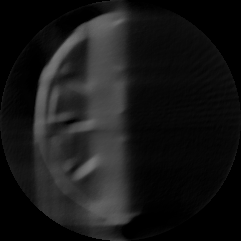}&
\includegraphics[align=c,trim={0 0.5cm 2cm 0.5cm},clip,width=0.2\textwidth]{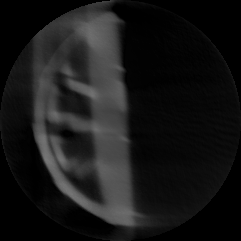}
\\
\\
\rotatebox[origin=c]{90}{\textbf{Multi-slice fusion}} & \rotatebox[origin=c]{90}{\textbf{(proposed)}} &
\includegraphics[align=c,trim={0 0.5cm 2cm 0.5cm},clip,width=0.2\textwidth]{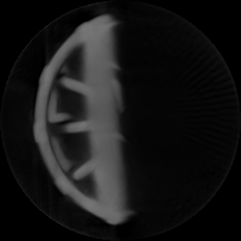}&
\includegraphics[align=c,trim={0 0.5cm 2cm 0.5cm},clip,width=0.2\textwidth]{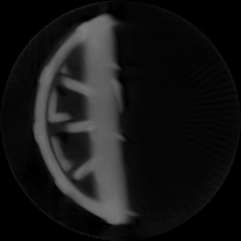}&
\includegraphics[align=c,trim={0 0.5cm 2cm 0.5cm},clip,width=0.2\textwidth]{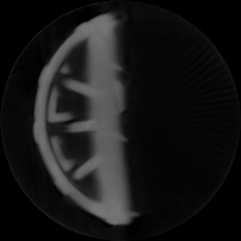}&
\includegraphics[align=c,trim={0 0.5cm 2cm 0.5cm},clip,width=0.2\textwidth]{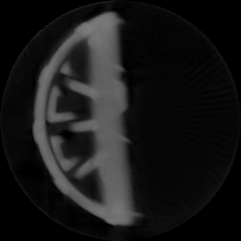}
\\
\\

\end{tabular}
\endgroup
\end{center}

\caption{ Comparison of different methods for simulated data with $90^{\circ}$ rotation of object per time-point.
The FBP reconstruction has severe limited-angle artifacts.
MBIR+TV improves the reconstruction in some regions but it suffers in areas affected by limited angular information. 
MBIR+4D-MRF reduces limited-angle artifacts, but allows severe artifacts to form that are not necessarily consistent with real 4D image sequences.
In contrast, the multi-slice fusion result does not suffer from major limited-angle artifacts.
}
\label{fig:simresults_90_xy}
\end{figure*}

\begin{figure}[!ht]
\centering     

\subfigure{\includegraphics[width=0.5\textwidth]{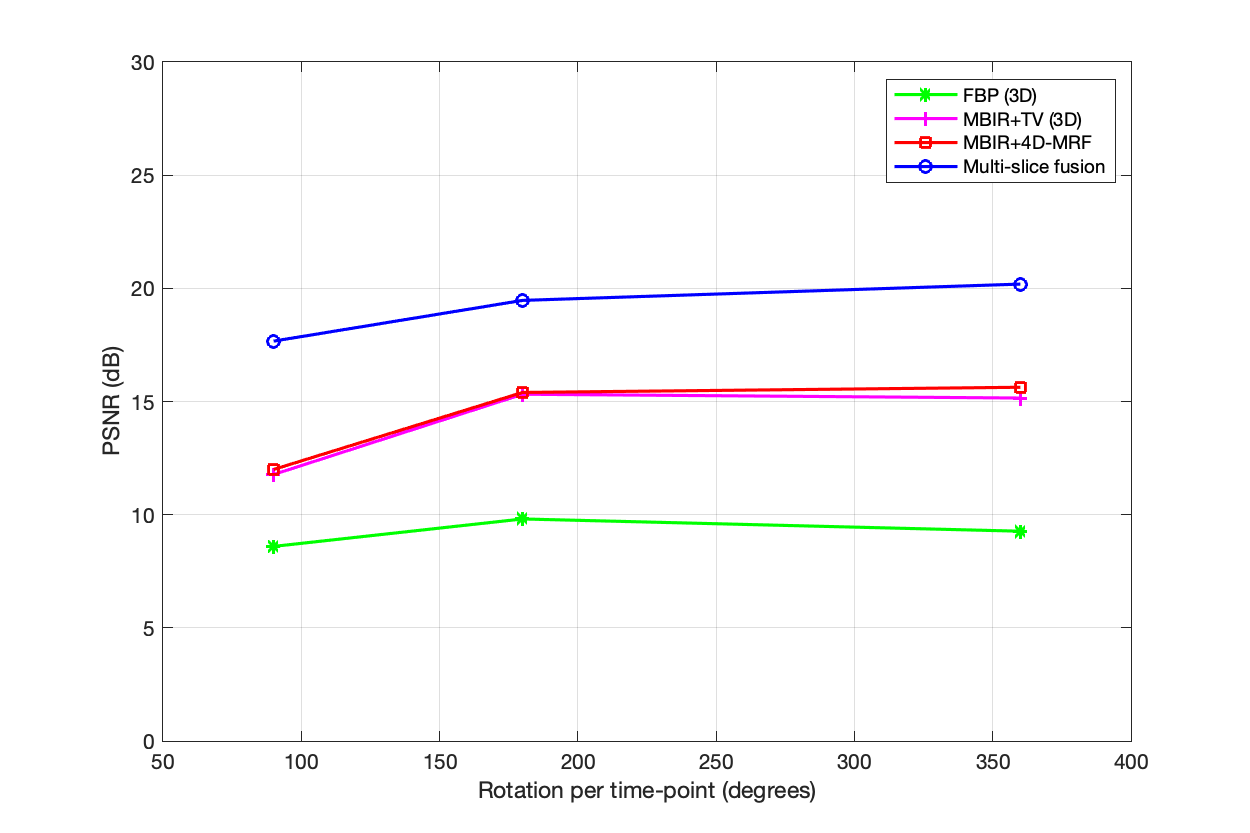}}

\begin{center}
\begingroup
\setlength{\tabcolsep}{1.2pt} 
\renewcommand{\arraystretch}{0.2} 
\begin{tabular}{cccccc}
&
$90^\circ$ & $180^\circ$ & $360^\circ$ \\
& (4 views) & (8 views) & (16 views) \\
\\
Phantom &
\includegraphics[align=c,trim={0cm 0cm 0cm 0},clip,width=0.1\textwidth]{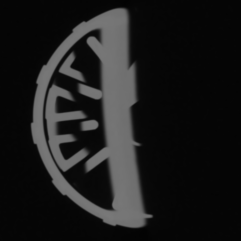}&
\includegraphics[align=c,trim={0cm 0cm 0cm 0},clip,width=0.1\textwidth]{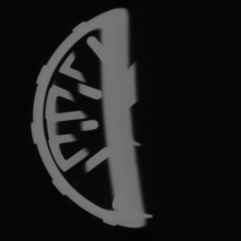}&
\includegraphics[align=c,trim={0cm 0cm 0cm 0},clip,width=0.1\textwidth]{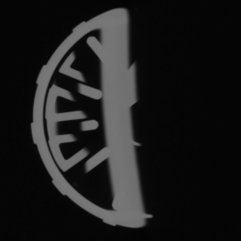}
\\
\\
FBP (3D) &
\includegraphics[align=c,trim={0cm 0cm 0cm 0},clip,width=0.1\textwidth]{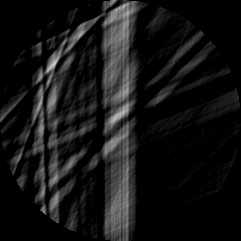}&
\includegraphics[align=c,trim={0cm 0cm 0cm 0},clip,width=0.1\textwidth]{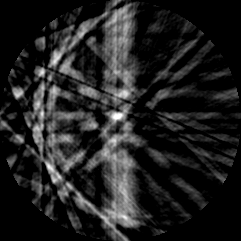}&
\includegraphics[align=c,trim={0cm 0cm 0cm 0},clip,width=0.1\textwidth]{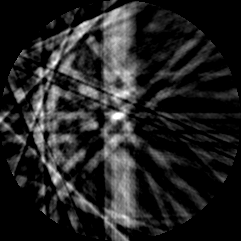}
\\
\\
MBIR+TV (3D) &
\includegraphics[align=c,trim={0cm 0cm 0cm 0},clip,width=0.1\textwidth]{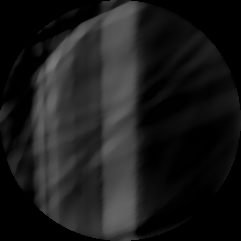}&
\includegraphics[align=c,trim={0cm 0cm 0cm 0},clip,width=0.1\textwidth]{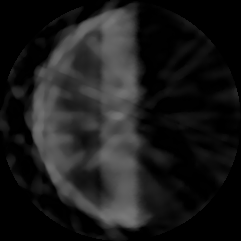}&
\includegraphics[align=c,trim={0cm 0cm 0cm 0},clip,width=0.1\textwidth]{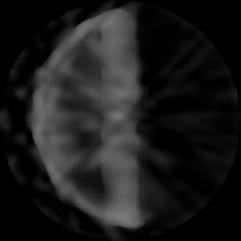}
\\
\\
MBIR+4D-MRF &
\includegraphics[align=c,trim={0cm 0cm 0cm 0},clip,width=0.1\textwidth]{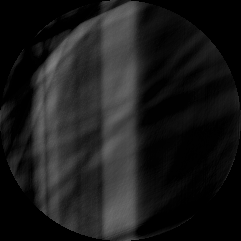}&
\includegraphics[align=c,trim={0cm 0cm 0cm 0},clip,width=0.1\textwidth]{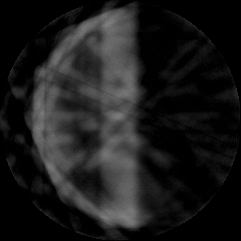}&
\includegraphics[align=c,trim={0cm 0cm 0cm 0},clip,width=0.1\textwidth]{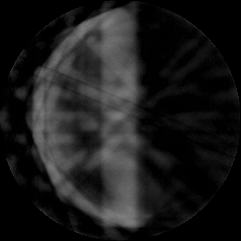}
\\
\\
\textbf{Multi-slice fusion} &
\includegraphics[align=c,trim={0cm 0cm 0cm 0},clip,width=0.1\textwidth]{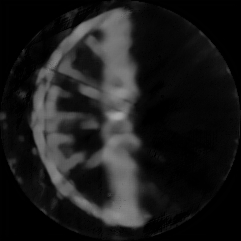}&
\includegraphics[align=c,trim={0cm 0cm 0cm 0},clip,width=0.1\textwidth]{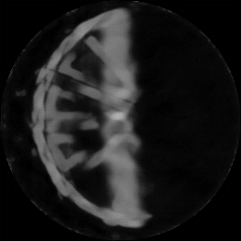}&
\includegraphics[align=c,trim={0cm 0cm 0cm 0},clip,width=0.1\textwidth]{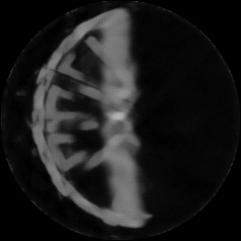}
\end{tabular}
\endgroup
\end{center}

\caption{Illustration of the reconstruction quality obtained for extreme sparse-view data with different levels of limited angle per time-point. 
FBP results in strong artifacts due to sparse-views and limited angles.
MBIR+TV and MBIR+4D-MRF mitigates most of the major sparse-view artifacts but suffers from limited angle artifacts in the $90^\circ$ limited angle case.
Multi-slice fusion results in fewer limited-angle and sparse-view artifacts and an improved PSNR metric.
Moreover, multi-slice fusion results in reduced artifacts compared to MBIR+TV and MBIR+4D-MRF as the rotation per time point is decreased.
}
\label{fig:results_lim_angle_vary}
\end{figure}

\begin{figure*}[!htb]
\centering     

\subfigure[FBP (3D)]{\includegraphics[trim={0.9cm 0 0 1.5cm},clip,width=0.32\textwidth]
{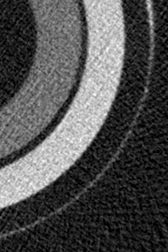}}
\subfigure[MBIR+4D-MRF]{\includegraphics[trim={0.9cm 0 0 1.5cm},clip,width=0.32\textwidth]
{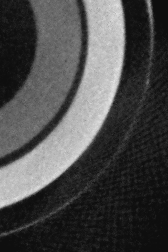}}
\subfigure[\textbf{Multi-slice fusion (proposed)}]
{
\begin{tikzpicture}
    \node[anchor=south west,inner sep=0] (image) at (0,0) {\includegraphics[trim={0.9cm 0 0 1.5cm},clip,width=0.32\textwidth]{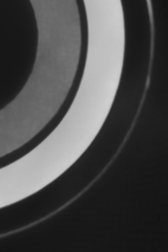}};
    \begin{scope}[x={(image.south east)},y={(image.north west)}]
        \draw[blue,ultra thick,rotate around={45:(0.5,0.35)}] (0.5,0.35) ellipse (0.5 and 0.07);
    \end{scope}
\end{tikzpicture}
}
\subfigure[MBIR+$H_\text{xy,t}$ (Missing Feature)]{\includegraphics[trim={0.9cm 0 0 1.5cm},clip,width=0.32\textwidth]
{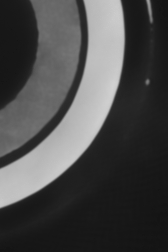}}
\subfigure[MBIR+$H_\text{yz,t}$ (Horizontal Streaks)]{\includegraphics[trim={0.9cm 0 0 1.5cm},clip,width=0.32\textwidth]
{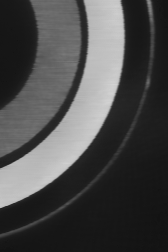}}
\subfigure[MBIR+$H_\text{zx,t}$ (Vertical Streaks)]
{
\begin{tikzpicture}
    \node[anchor=south west,inner sep=0] (image) at (0,0) {\includegraphics[trim={0.9cm 0 0 1.5cm},clip,width=0.32\textwidth]{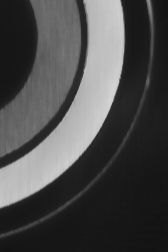}};
\end{tikzpicture}
}
\caption{ Comparison of different methods for Real Data $360^\circ$: vial.
Each image is a slice through the reconstructed vial for one time-point along the spatial xy-plane.
Both FBP and MBIR+4D-MRF suffer from obvious windmill artifacts, higher noise and blurred edges.
In contrast to that, the multi-slice fusion reconstruction has smooth and uniform textures while preserving edge definition.
MBIR+$H_\text{yz,t}$ and MBIR+$H_\text{zx,t}$ suffer from horizontal and vertical streaks.
MBIR+$H_\text{xy,t}$ cannot reconstruct the outer ring since the slice displayed is at the edge of the aluminum seal and the xy-plane does not contain sufficient information.
Multi-slice fusion can resolve the edges of the rings better than either of MBIR+$H_\text{xy,t}$, MBIR+$H_\text{yz,t}$, and MBIR+$H_\text{zx,t}$ since it has information from all the spatial coordinates.
}
\label{fig:results_xy}
\end{figure*}

\begin{figure}[!htb]
\centering     
\subfigure[FBP (3D)]{\includegraphics[width=0.15\textwidth]{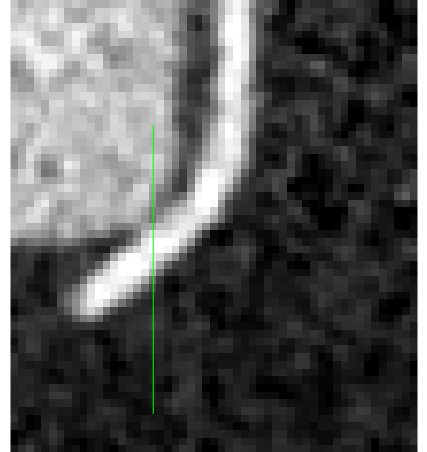}}
\subfigure[MBIR+4D-MRF]{\includegraphics[width=0.15\textwidth]{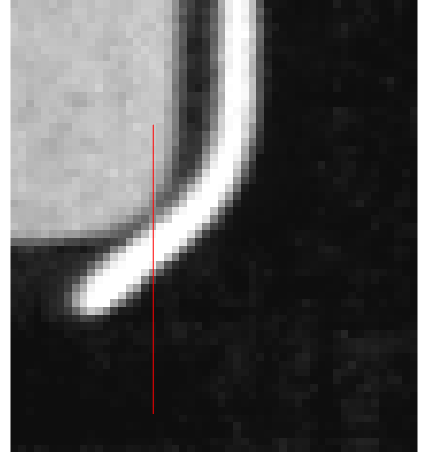}}
\subfigure[\textbf{Multi-slice fusion \newline (proposed)}]{\includegraphics[width=0.15\textwidth]{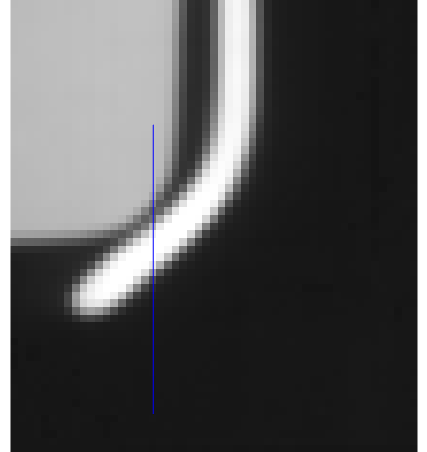}}
\subfigure[Plot of cross-section]{\includegraphics[width=0.5\textwidth]{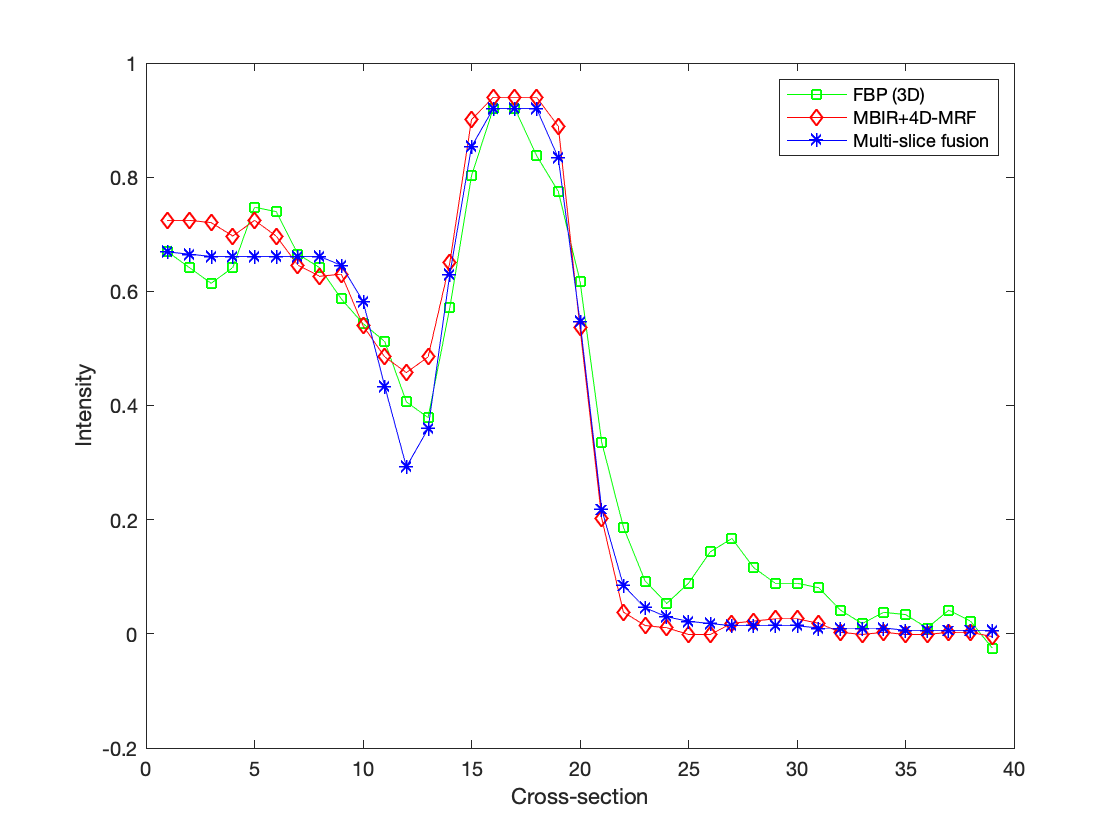}}
\caption{ Plot of cross-section through the vial at a time when the aluminum and glass have physically separated.
Multi-slice fusion is able to resolve the junction between materials better while simultaneously producing a smoother reconstruction within materials compared to MBIR+4D-MRF and FBP.}
\label{fig:results_crossSection}
\end{figure} 

\begin{figure}[!htb]
\centering     
\includegraphics[width=0.5\textwidth]{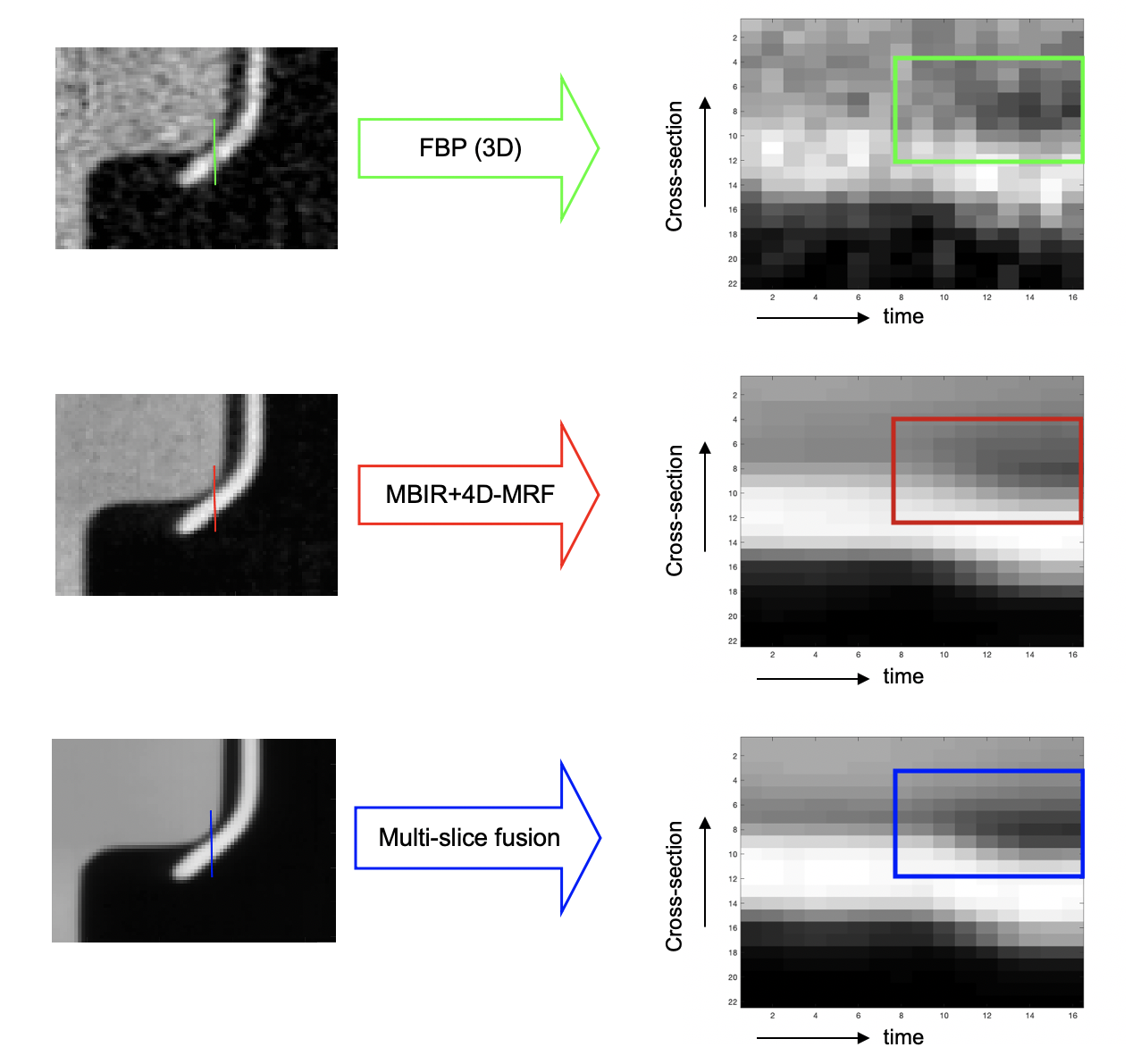}
\caption{Illustration of temporal resolution for real data $360^\circ$ : vial.
We plot a cross-section through the vial with time for each method: multi-slice fusion, MBIR+4D-MRF, FBP.
Multi-slice fusion results in improved space-time resolution of the separation of aluminum and glass.}
\label{fig:time_resolution}
\end{figure}

\begin{figure*}[!ht]
\centering     

\begin{center}
\begingroup
\setlength{\tabcolsep}{1.2pt} 
\renewcommand{\arraystretch}{0.2} 
\begin{tabular}{cccccc}
& &
Time-point 1 & Time-point 2 & Time-point 3 & Time-point 4\\
\\
\rotatebox[origin=c]{90}{FBP (3D)} & &
\includegraphics[align=c,trim={2cm 0cm 0cm 0},clip,width=0.21\textwidth]{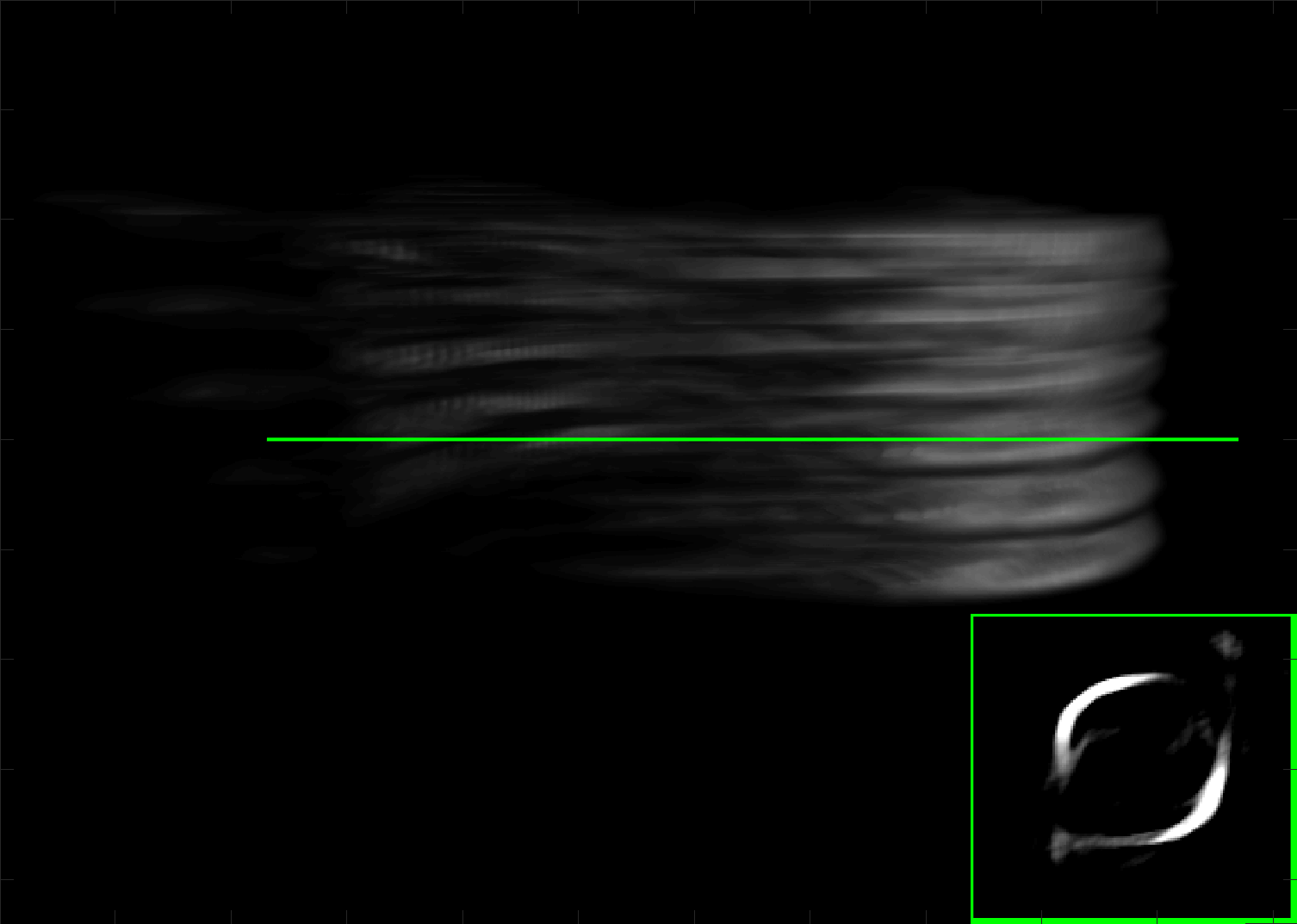}&
\includegraphics[align=c,trim={2cm 0cm 0cm 0},clip,width=0.21\textwidth]{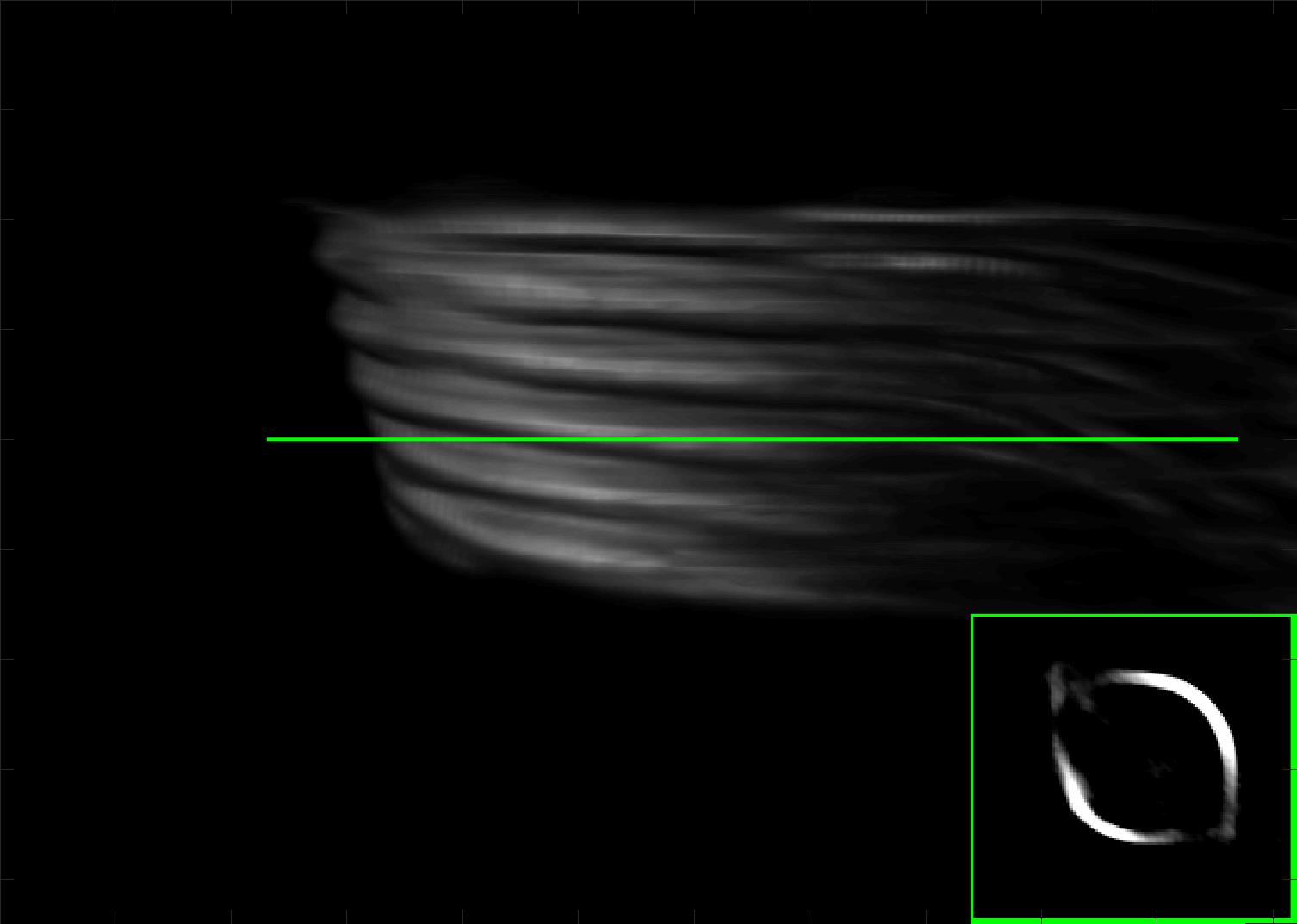}&
\includegraphics[align=c,trim={2cm 0cm 0cm 0},clip,width=0.21\textwidth]{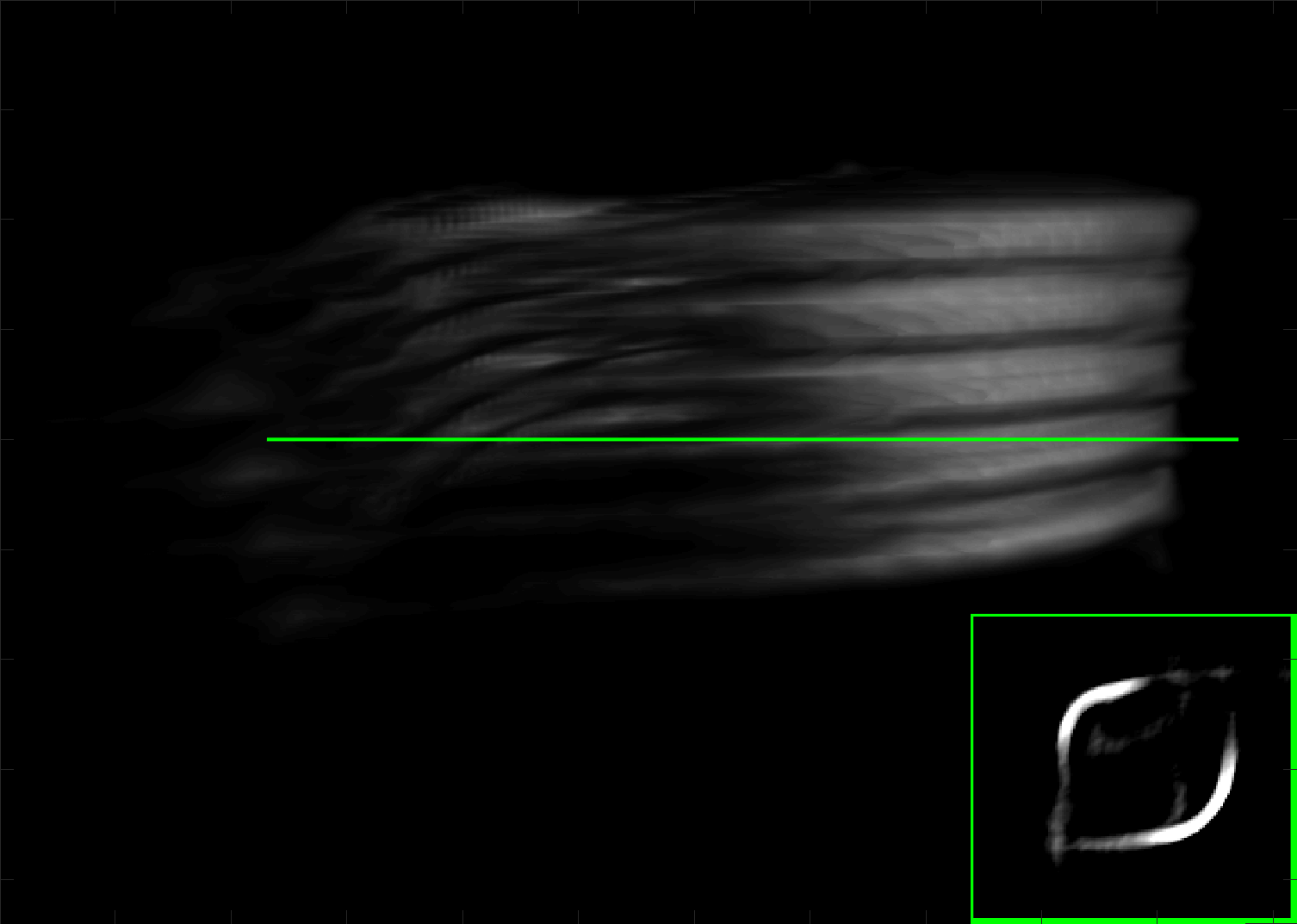}&
\includegraphics[align=c,trim={2cm 0cm 0cm 0},clip,width=0.21\textwidth]{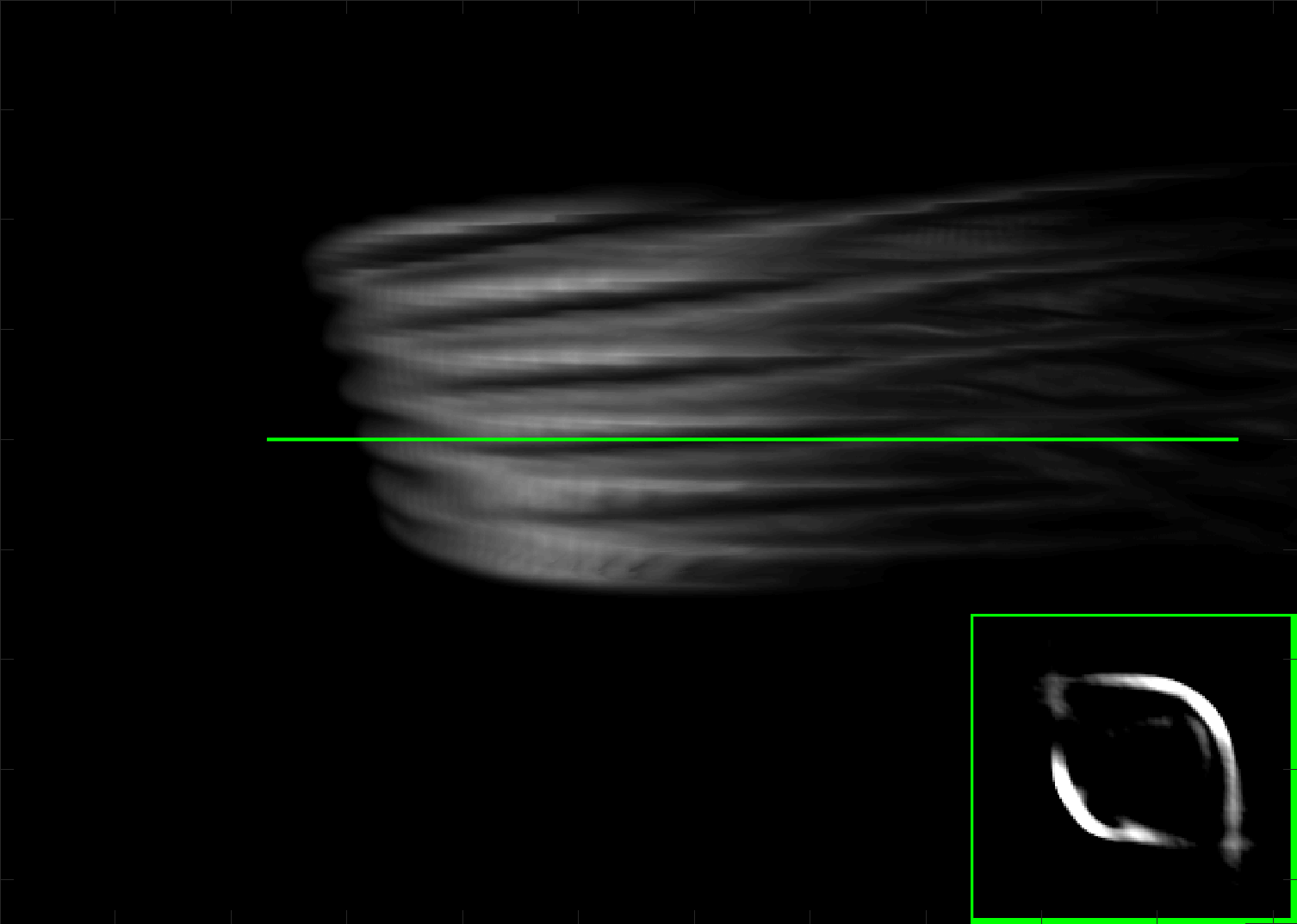}
\\
\\
\rotatebox[origin=c]{90}{MBIR+4D-MRF} & &
\includegraphics[align=c,trim={2cm 0cm 0cm 0},clip,width=0.21\textwidth]{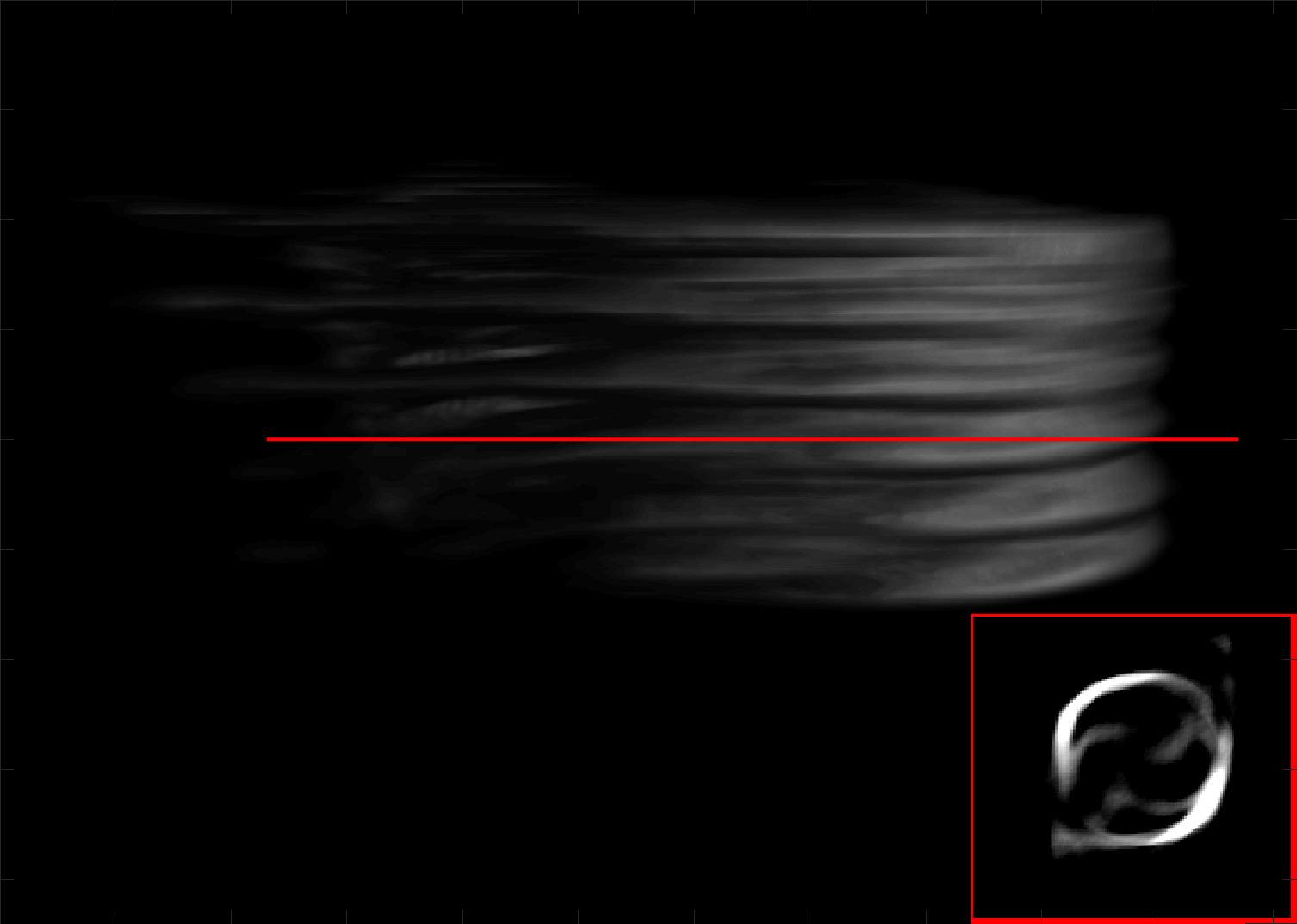}&
\includegraphics[align=c,trim={2cm 0cm 0cm 0},clip,width=0.21\textwidth]{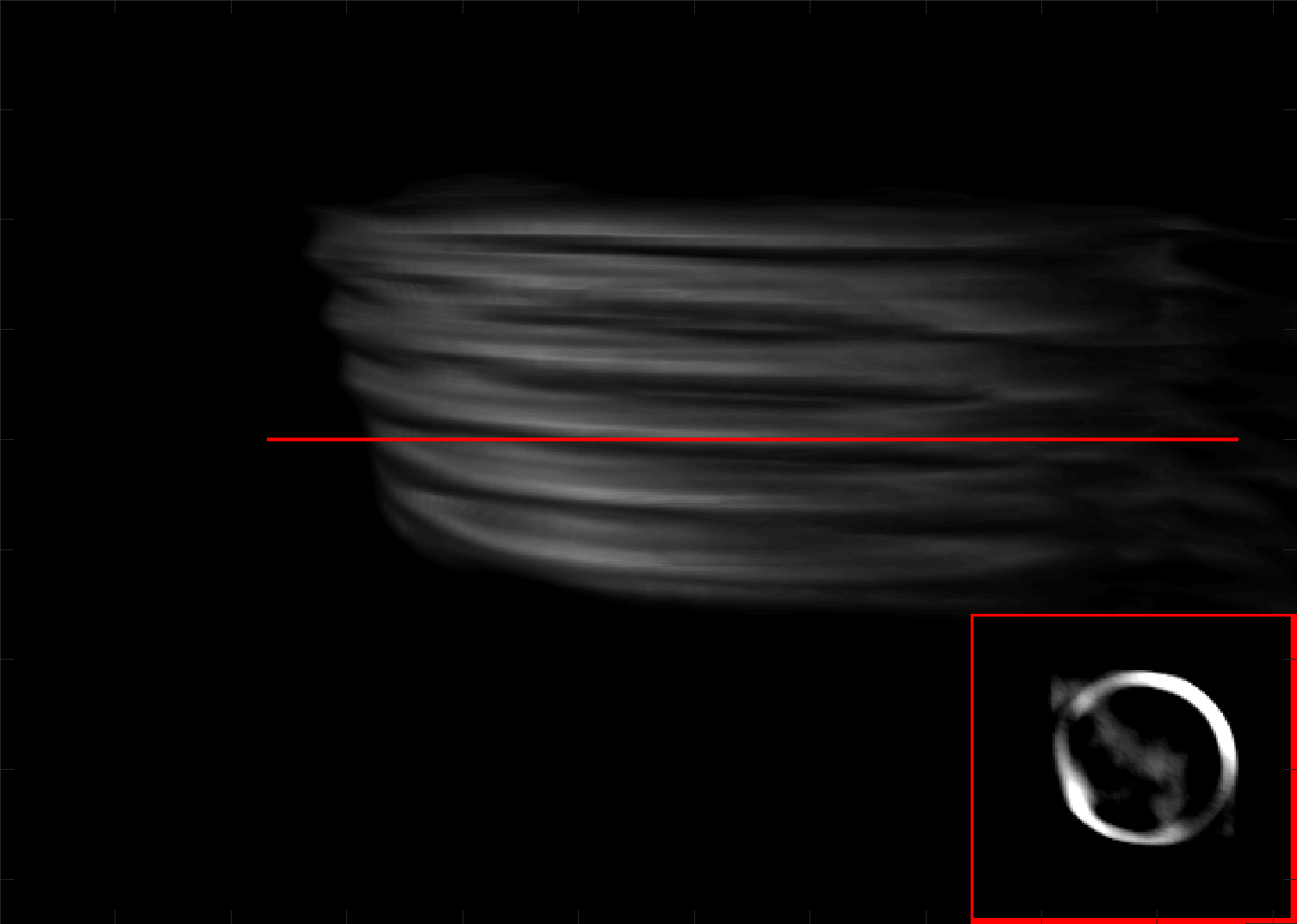}&
\includegraphics[align=c,trim={2cm 0cm 0cm 0},clip,width=0.21\textwidth]{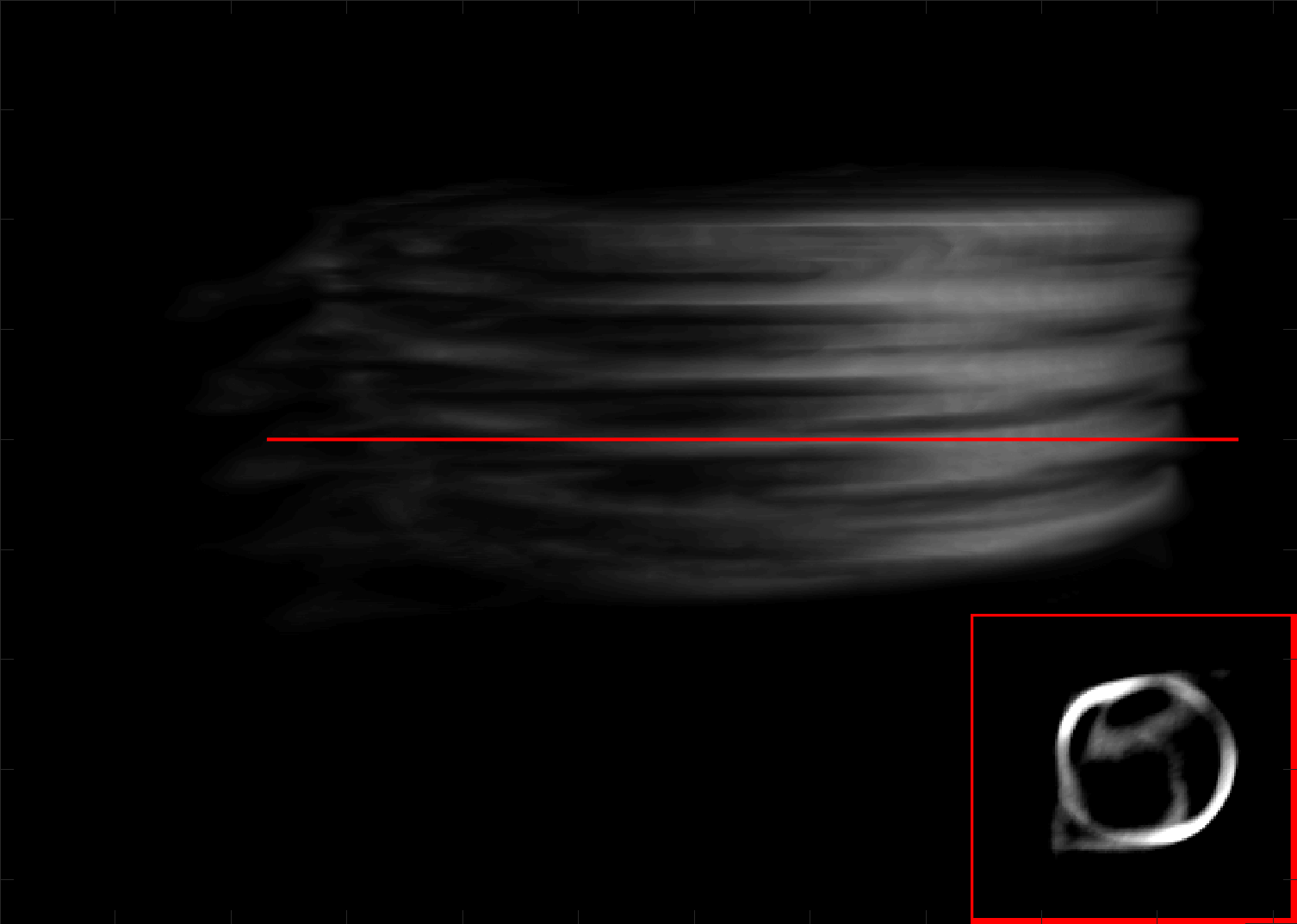}&
\includegraphics[align=c,trim={2cm 0cm 0cm 0},clip,width=0.21\textwidth]{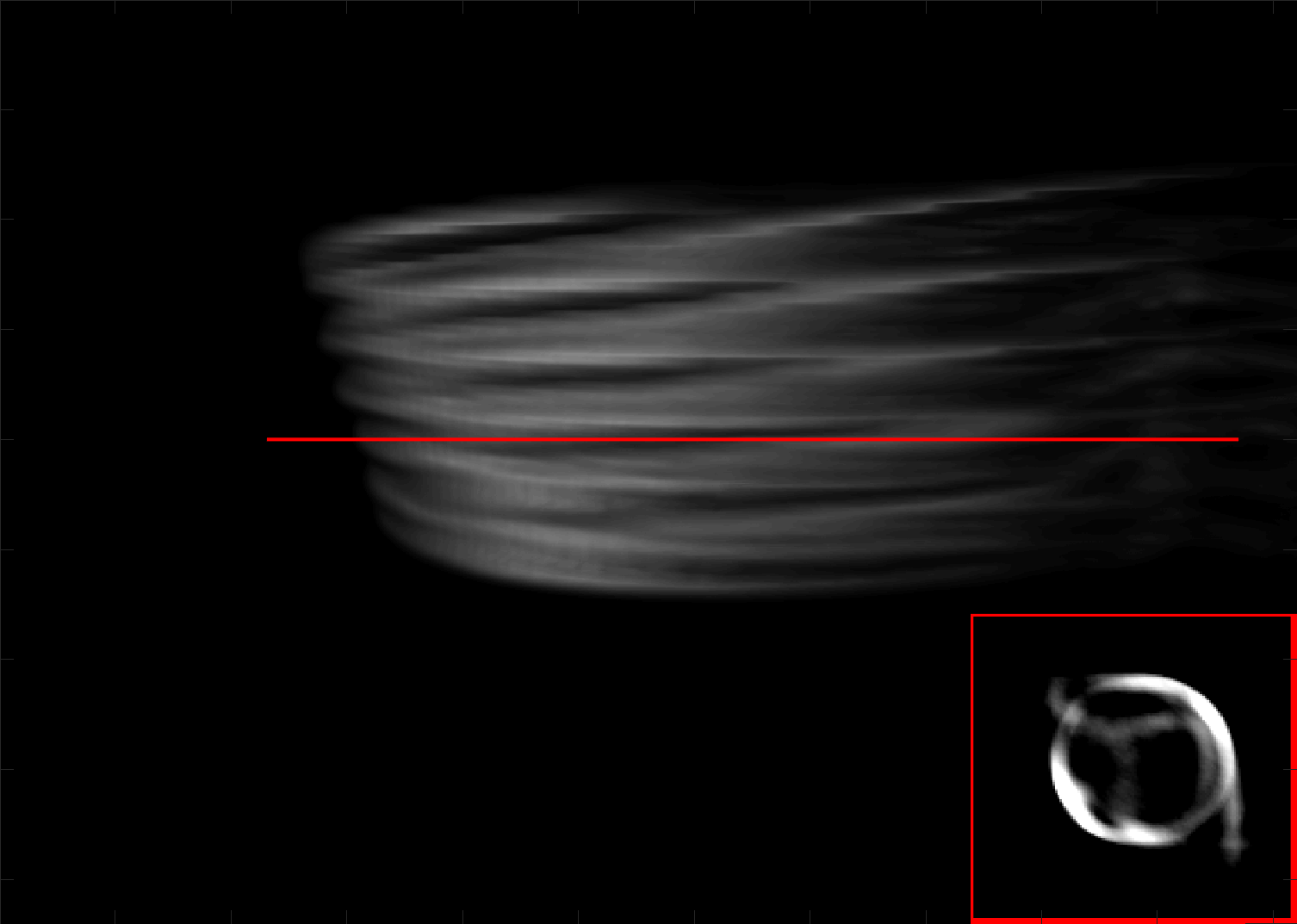}
\\
\\
\rotatebox[origin=c]{90}{\textbf{Multi-slice fusion}} & \rotatebox[origin=c]{90}{\textbf{(proposed)}} &
\includegraphics[align=c,trim={2cm 0cm 0cm 0},clip,width=0.21\textwidth]{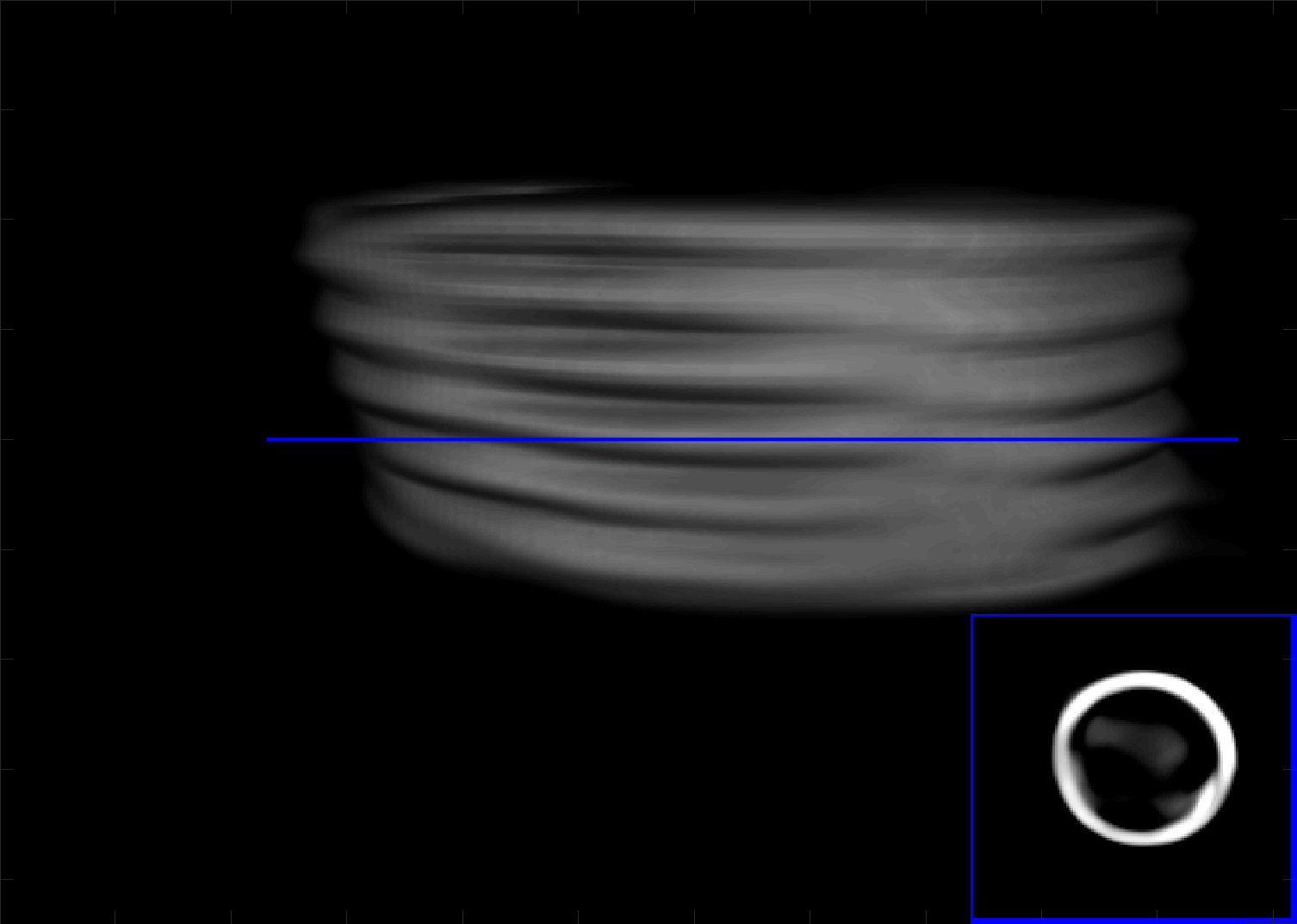}&
\includegraphics[align=c,trim={2cm 0cm 0cm 0},clip,width=0.21\textwidth]{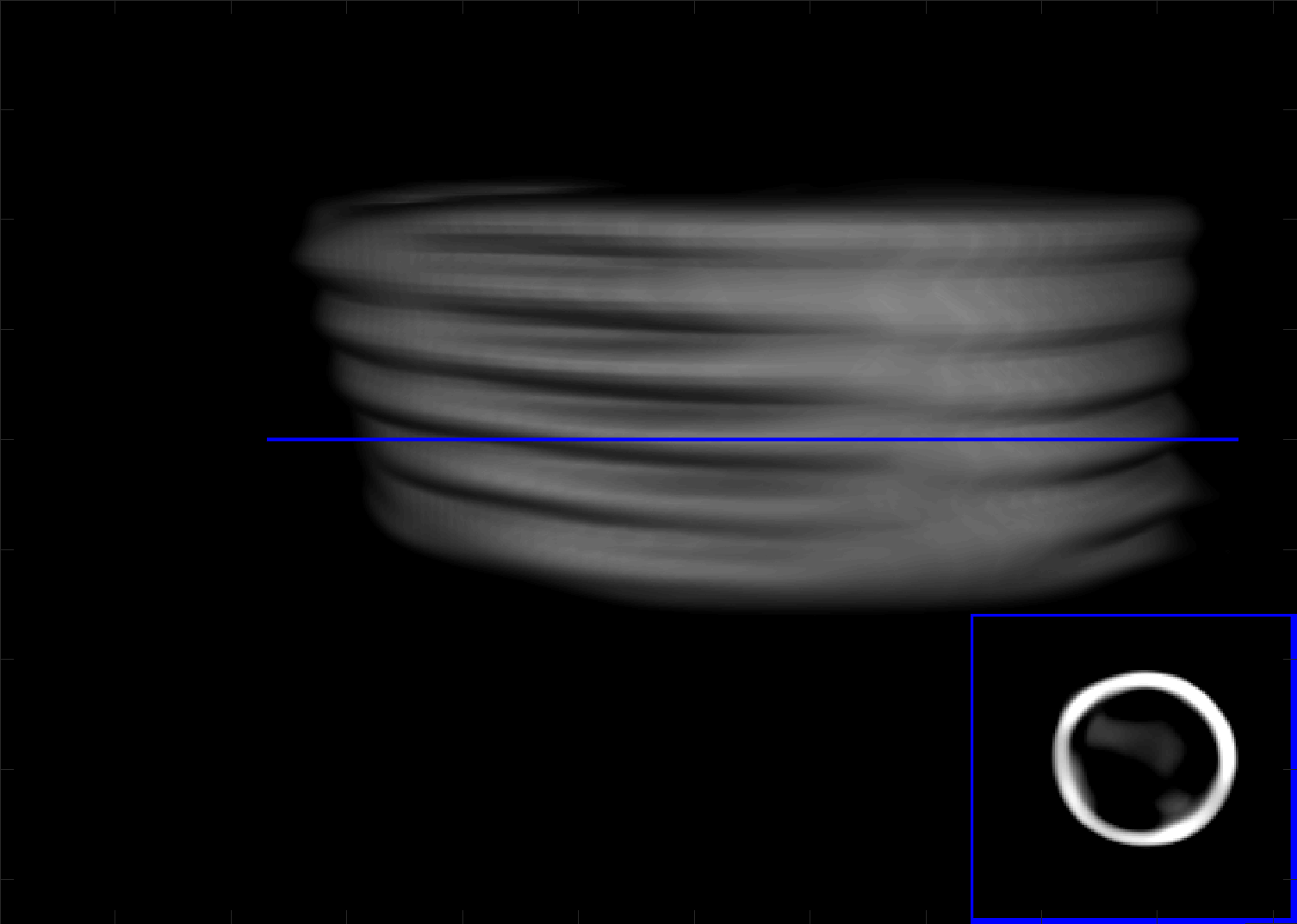}&
\includegraphics[align=c,trim={2cm 0cm 0cm 0},clip,width=0.21\textwidth]{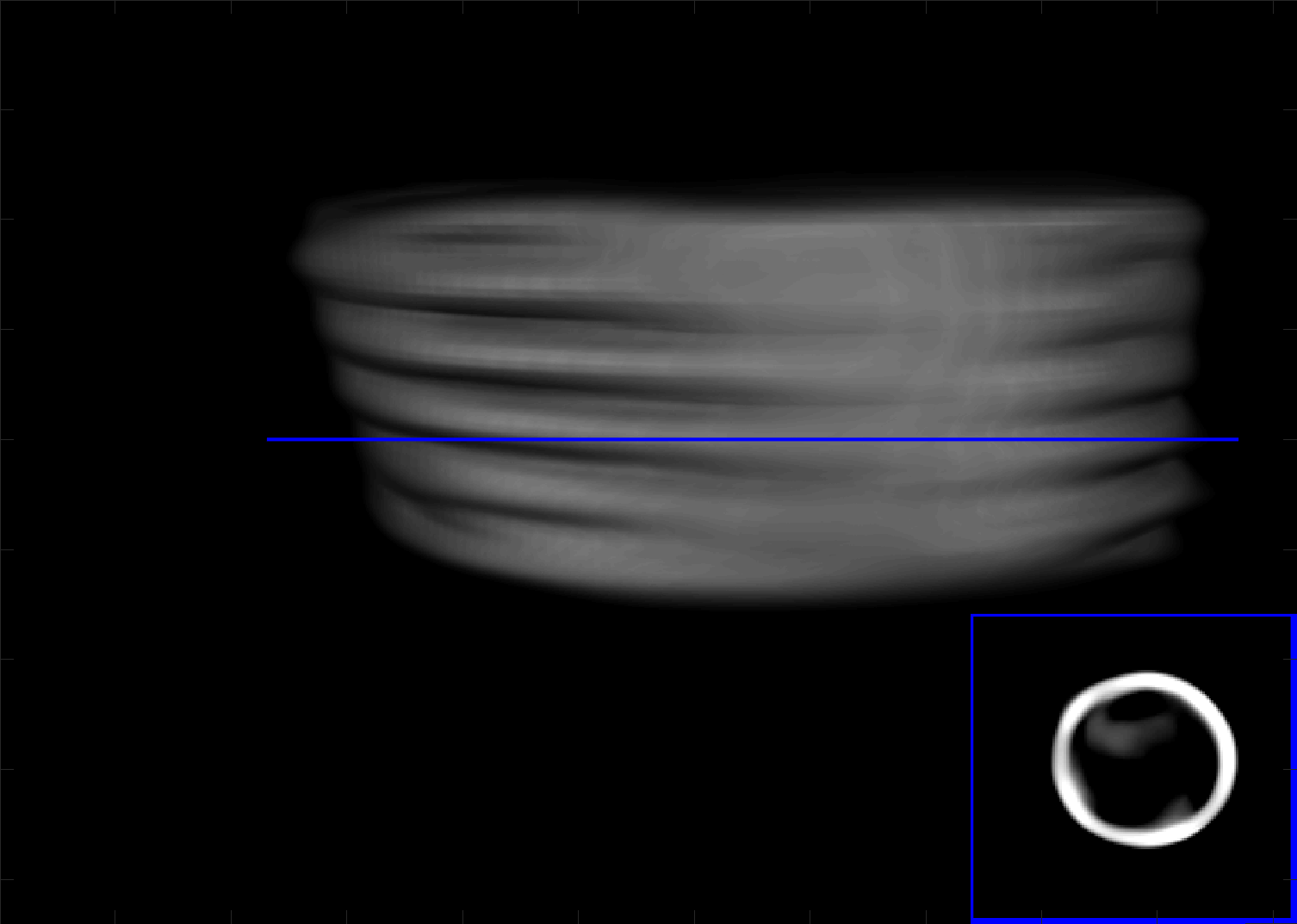}&
\includegraphics[align=c,trim={2cm 0cm 0cm 0},clip,width=0.21\textwidth]{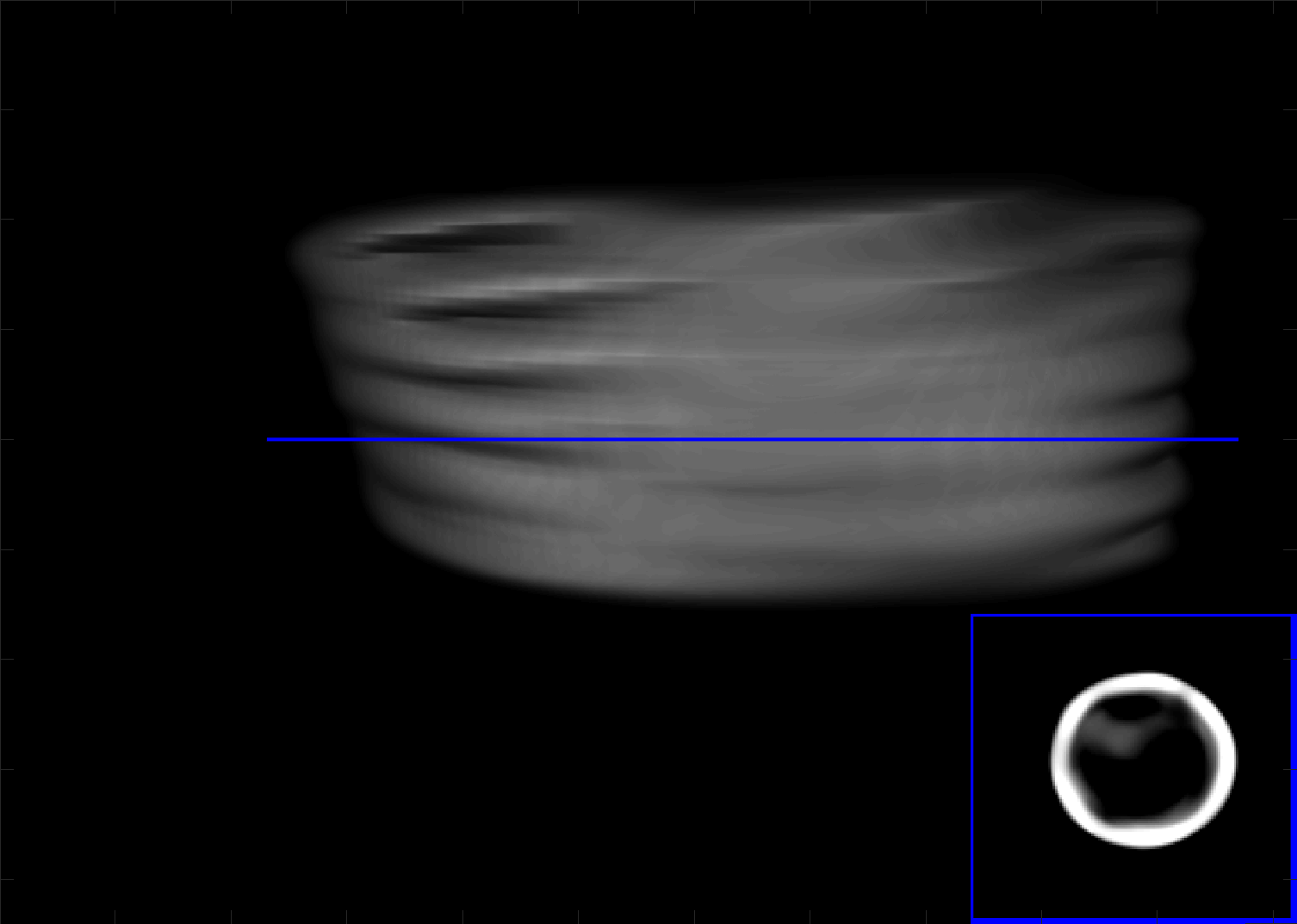}
\end{tabular}
\endgroup
\end{center}

\caption{Volume rendering of the reconstructed spring and its cross-section for four time-points.
A $90^{\circ}$ limited set of views is used to reconstruct each time-point. 
The FBP reconstruction contains severe limited-angle artifacts.
MBIR+4D-MRF mitigates some limited-angle artifacts but some artifacts remain.
}
\label{fig:results_kwikpen_xy}
\end{figure*}


\subsection{Simulated Data $360^\circ$}
\label{sec:sim360}

\begin{table}[!htb]
\centering{} 
\small
\begin{tabular}{r|l}
\toprule
Magnification & 5.57 \\
Number of Views per Time-point & 75 \\
Rotation per Time-point & $360^{\circ}$ \\
Cropped Detector Array & $240 \times 28$ , $(0.95 \ \mathrm{mm})^2$ \\
Voxel Size & $(0.17 \ \mathrm{mm})^3$ \\
Reconstruction Size (x,y,z,t) & $240 \times 240 \times 28 \times 8 $ \\
\bottomrule
\hline
\end{tabular}
\\
\vspace{1mm}
\caption{Experimental specifications for Simulated Data $360^\circ$}
\label{table:simsetup_360}
\end{table}

In this section we present results on simulated data to evaluate our method in a sparse-view setting.
Each time-point is reconstructed from a sparse set of views spanning $360^\circ$.
We take a low-noise CT reconstruction of a bottle and screw cap and denoise it further using BM4D~\cite{bm4d} to generate a clean 3D volume to be used as a 3D phantom.
We then vertically translate the 3D phantom by one pixel per time-point to generate a 4D phantom $x^0$.
We generate simulated sinogram measurements as $\mathcal{N}(Ax^0, \Lambda^{-1})$ where $A$ is the projection matrix and the inverse covariance matrix $\Lambda=\text{diag}\{ c \exp \left\{ - Ax^0 \right\}$ accounts for the non-uniform noise variance due to a Gaussian approximation~\cite{bouman1996unified} of the underlying Poisson noise.
We then perform a 4D reconstruction from the simulated sinogram data and compare with the 4D phantom.
The experimental specifications are summarized in Table~\ref{table:simsetup_360}.

Figure~\ref{fig:simresults_360_xy} compares reconstructions using multi-slice fusion with several other methods.
Each image is a slice through the reconstructed object for one time-point along the spatial xy-plane.
The reconstruction using FBP suffers from high noise and fails to recover the small hole in the bottom of the image.
The reconstructions using MBIR+TV and MBIR+4D-MRF suffer from jagged edges and fail to recover the small hole in the bottom of the image.
MBIR+$H_\text{yz,t}$ and MBIR+$H_\text{zx,t}$ suffer from horizontal and vertical streaks, respectively, since the denoisers were applied in those planes.
MBIR+$H_\text{xy,t}$ does not suffer from streaks in the figure since we are viewing a slice along the xy-plane, but it suffers from other artifacts.
MBIR+$H_\text{xy,t}$ cannot reconstruct the small hole in the bottom of the image since the xy-plane does not contain sufficient information.
It is to be noted that multi-slice fusion enhances the size and contrast of the small hole highlighted by the blue circle relative to the phantom.
This can cause deviations when measuring the size of small features in the reconstruction.

Next we plot a cross-section through the object for multi-slice fusion, MBIR+4D-MRF, MBIR+TV, FBP, and the phantom in Figure~\ref{fig:simresults_360_crossSection}.
Multi-slice fusion results in the most accurate reconstruction of the gap between materials.

Finally we report the peak signal to noise ratio (PSNR) and the structural similarity index measure (SSIM)~\cite{SSIM} with respect to the phantom for each method in Table~\ref{table:sim_360_metrics} to objectively measure image quality.
We define the PSNR for a given 4D reconstruction $x$ with a phantom $x^0$ as
\begin{equation}
    \text{PSNR}(x) = 20 \log_{10} \left( \dfrac{\text{Range}(x^0)}{\text{RMSE}(x,x^0)}
    \right) ,
\end{equation}
where range is computed from the $0.1^{\text{st}}$ and $99.9^{\text{th}}$ percentiles of the phantom.
As can be seen from Table~\ref{table:sim_360_metrics}, multi-slice fusion results in the highest PSNR and SSIM scores.

\begin{table}[!htb]
\centering{} 
\small
\begin{tabular}{r|l|l}
\toprule
Method & PSNR(dB) & SSIM\\
\hline
FBP & 19.69 & 0.609 \\
MBIR+TV & 26.63 & 0.860 \\
MBIR+4D-MRF & 25.84 & 0.787 \\
Multi-slice fusion & \textbf{29.07} & \textbf{0.943} \\
MBIR+$H_\text{xy,t}$ & 29.03 & 0.922 \\
MBIR+$H_\text{yz,t}$ & 28.04 & 0.932 \\
MBIR+$H_\text{zx,t}$ & 28.31 & 0.926 \\
\bottomrule
\hline
\end{tabular}
\\
\vspace{1mm}
\caption{Quantitative Evaluation for simulated data $360^\circ$.
Multi-slice fusion has the highest PSNR and SSIM metric among all the methods.}
\label{table:sim_360_metrics}
\end{table}

\subsection{Simulated Data $90^\circ$}

\begin{table}[!htb]
\centering{} 
\small
\begin{tabular}{r|l}
\toprule
Magnification & 5.57 \\
Number of Views per Time-point & 36 \\
Rotation per Time-point & $90^{\circ}$ \\
Cropped Detector Array & $240 \times 28$ , $(0.95 \ \mathrm{mm})^2$ \\
Voxel Size & $(0.17 \ \mathrm{mm})^3$ \\
Reconstruction Size (x,y,z,t) & $240 \times 240 \times 28 \times 8 $ \\
\bottomrule
\hline
\end{tabular}
\\
\vspace{1mm}
\caption{Experimental specifications for Simulated Data $90^\circ$}
\label{table:simsetup_90}
\end{table}

In this section we present results on simulated data to evaluate our method in a sparse-view and limited-angle setting.
Each time-point is reconstructed from a sparse set of views spanning $90^\circ$.
The simulated measurement data is generated in a similar fashion as Section~\ref{sec:sim360} using the experimental specifications summarized in Table~\ref{table:simsetup_90}.

Figure~\ref{fig:simresults_90_xy} shows a comparison of different methods for simulated data with $90^{\circ}$ rotation of object per time-point.
The FBP reconstruction has severe limited-angle artifacts.
MBIR+TV improves the reconstruction in some regions but it suffers in areas affected by limited angular information. 
MBIR+4D-MRF reduces limited-angle artifacts, but allows severe artifacts to form that are not necessarily consistent with real 4D image sequences.
In contrast, the multi-slice fusion result does not suffer from major limited-angle artifacts.

Table~\ref{table:sim_90_metrics} shows peak signal to noise ratio (PSNR) and structural similarity index measure (SSIM) with respect to the phantom for each method.
Multi-slice fusion results in the highest PSNR and SSIM scores.

\begin{table}[!htb]
\centering{} 
\small
\begin{tabular}{r|l|l}
\toprule
Method & PSNR(dB) & SSIM\\
\hline
FBP & 10.86 & 0.467 \\
MBIR+TV & 15.35 & 0.801 \\
MBIR+4D-MRF & 14.25 & 0.742 \\
Multi-slice fusion & \textbf{19.44} & \textbf{0.875} \\
\bottomrule
\hline
\end{tabular}
\\
\vspace{1mm}
\caption{Quantitative Evaluation for simulated data $90^\circ$.
Multi-slice fusion has the highest PSNR and SSIM metric among all the methods.}
\label{table:sim_90_metrics}
\end{table}

In order to determine the effectiveness of our method for more challenging data, we generate extreme sparse-view simulated data with different angle of rotation per time-point while keeping the rest of the experimental setup the same as Table~\ref{table:simsetup_90}. 
Figure~\ref{fig:results_lim_angle_vary} illustrates the reconstruction quality obtained for the extreme sparse-view data with different levels of limited angle. 
FBP results in strong artifacts due to sparse-views and limited angles.
MBIR+TV and MBIR+4D-MRF mitigates most of the major sparse-view artifacts but suffers from limited angle artifacts in the $90^\circ$ limited angle case.
Multi-slice fusion results in fewer limited-angle and sparse-view artifacts and an improved PSNR metric.
Moreover, multi-slice fusion results in a reduced motion and sparse view artifacts as compared to MBIR+TV and MBIR+4D-MRF as the rotation per time point is decreased.

\subsection{Real Data $360^\circ$: Vial Compression}

\begin{table}[!htb]
\centering{} 
\small
\begin{tabular}{r|l}
\toprule
Scanner Model & North Star Imaging X50 \\
Voltage & $140$ $\mathrm{kV}$ \\
Current & $500$ $\mathrm{\mu A}$ \\
Exposure & $20$ $\mathrm{ms}$ \\
Source-Detector Distance & $839$ $\mathrm{mm}$  \\
Magnification & $5.57$ \\
Number of Views per Time-point & $150$ \\
Rotation per Time-point & $360^{\circ}$ \\
Cropped Detector Array & $731 \times 91$, $(0.25 \ \mathrm{mm})^2$\\
Voxel Size & $(0.0456 \ \mathrm{mm})^3$ \\
Reconstruction Size (x,y,z,t) & $731 \times 731 \times 91 \times 16 $ \\
\bottomrule
\hline
\end{tabular}
\\
\vspace{1mm}
\caption{Experimental specifications for Real Data $360^\circ$: Vial Compression}
\label{table:setup_vial}
\end{table}

\begin{figure}[!htb]
\centering     
\includegraphics[align=c,trim={0cm 7cm 2cm 0},clip,width=0.4\textwidth]{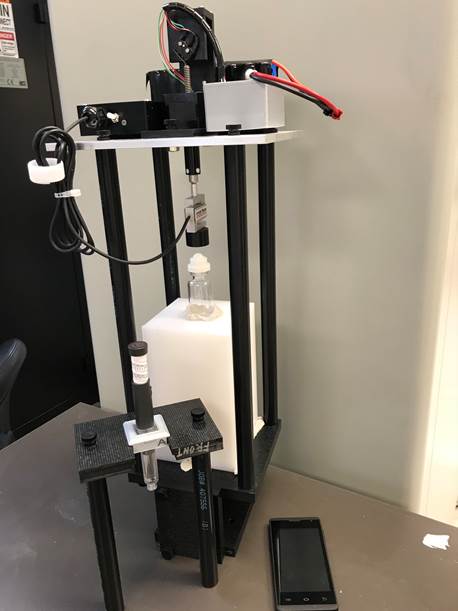}
\caption{Experimental setup for Real Data $360^\circ$: Vial Compression.
The vial is undergoing dynamic compression during the scan, to capture the mechanical response of the components.
The glass vial (center) and the actuator (top) are held together by a frame constructed of tubes and plates.
The tubes were placed outside the field of view of the CT scanner, thus causing artifacts in the reconstruction.
We describe a correction for this in Appendix~\ref{sec:jigcorr}.
}
\label{fig:vial_setup}
\end{figure}

In this section we present results on real data to evaluate our method in a sparse-view setting.
The data is from a dynamic cone-beam X-ray scan of a glass vial, with elastomeric stopper and aluminum crimp-seal, using a North Star Imaging X50 X-ray CT system.
The experimental specifications are summarized in Table~\ref{table:setup_vial}.

The vial is undergoing dynamic compression during the scan, to capture the mechanical response of the components as shown in Figure~\ref{fig:vial_setup}.
Of particular interest is the moment when the aluminum seal is no longer in contact with the underside of the glass neck finish.
This indicates the moment when the force applied exceeds that exerted by the rubber on the glass; this is known as the ``residual seal force'' \cite{vialCap}.

During the scan, the vial was held in place by fixtures that were placed out of the field of view as shown in Figure~\ref{fig:vial_setup}.
As the object rotated, the fixtures periodically intercepted the path of the X-rays resulting in corrupted measurements and consequently artifacts in the reconstruction. 
To mitigate this problem, we incorporate additional corrections that are described in Appendix~\ref{sec:jigcorr}.

Figure~\ref{fig:results_xy} compares multi-slice fusion with several other methods.
Each image is a slice through the reconstructed vial for one time-point along the spatial xy-plane.
Both FBP and MBIR+4D-MRF suffer from obvious artifacts, higher noise and blurred edges.
In contrast to that, the multi-slice fusion reconstruction has smooth and uniform textures while preserving edge definition.
Figure~\ref{fig:results_xy} also illustrates the effect of model fusion by comparing multi-slice fusion with MBIR+$H_\text{xy,t}$, MBIR+$H_\text{yz,t}$, and MBIR+$H_\text{zx,t}$.
MBIR+$H_\text{yz,t}$ and MBIR+$H_\text{zx,t}$ suffer from horizontal and vertical streaks respectively since the denoisers were applied in those planes.
MBIR+$H_\text{xy,t}$ does not suffer from streaks in the figure since we are viewing a slice along the xy-plane, but it suffers from other artifacts.
MBIR+$H_\text{xy,t}$ cannot reconstruct the outer ring since the slice displayed is at the edge of the aluminum seal and the xy-plane does not contain sufficient information.
In contrast, multi-slice fusion can resolve the edges of the rings better than either of MBIR+$H_\text{xy,t}$, MBIR+$H_\text{yz,t}$, and MBIR+$H_\text{zx,t}$ since it uses information from all the spatial coordinates.

Next, we plot a cross-section through the object for multi-slice fusion, MBIR+4D-MRF and FBP in Figure~\ref{fig:results_crossSection}. 
For this, we choose a time-point where we know the aluminum and glass have separated spatially, thus creating an air-gap.
Multi-slice fusion results in a deeper and more defined reconstruction of the gap between materials.
This supports that multi-slice fusion is able to preserve fine details in spite of producing a smooth regularized image.

Finally in Figure~\ref{fig:time_resolution} we plot a cross-section through the object with respect to time to show the improved space-time resolution of our method.
We do this for FBP, MBIR+4D-MRF and multi-slice fusion.
Multi-slice fusion results in improved space-time resolution of the separation of aluminum and glass.

\subsection{Real Data $90^\circ$: Injector Pen}

\begin{table}[!htb]
\centering{} 
\small
\begin{tabular}{r|l}
\toprule
Scanner Model & North Star Imaging X50 \\
Voltage & $165$ $\mathrm{kV}$ \\
Current & $550$ $\mathrm{\mu A}$ \\
Exposure & $12.5$ $\mathrm{ms}$ \\
Source-Detector Distance & $694$ $\mathrm{mm}$  \\
Magnification & $2.83$ \\
Number of Views per Time-point & $36$ \\
Rotation per Time-point & $90^{\circ}$ \\
Cropped Detector Array & $263 \times 768$, $(0.254 \ \mathrm{mm})^2$\\
Voxel Size & $(0.089 \ \mathrm{mm})^3$ \\
Reconstruction Size (x,y,z,t) & $263 \times 263 \times 778 \times 12 $ \\
\bottomrule
\hline
\end{tabular}
\\
\vspace{1mm}
\caption{Experimental specifications for Real Data $90^\circ$: Injector Pen}
\label{table:setup_kwikpen}
\end{table}


In this section we present results on real data to evaluate our method in a sparse-view and limited-angle setting.
The data is from a dynamic cone-beam X-ray scan of an injector pen using a North Star Imaging X50 X-ray CT system.
The experimental specifications are summarized in Table~\ref{table:setup_kwikpen}.

The injection device is initiated before the dynamic scan starts and completes a full injection during the duration of the scan.
We are interested in observing the motion of a particular spring within the injector pen in order to determine whether it is working as expected.
The spring in question is a non-helical wave-spring~\cite{komura1997wave} that is constructed out of circular rings that are joined together.
The spring exhibits a fast motion and as a result we need a high temporal resolution to observe the motion of the spring.
To have sufficient temporal resolution we reconstruct one frame for every $90^{\circ}$ rotation of the object instead of the conventional $360^{\circ}$ rotation.

\begin{figure*}[ht]
\centering     
\includegraphics[width=0.8\textwidth]{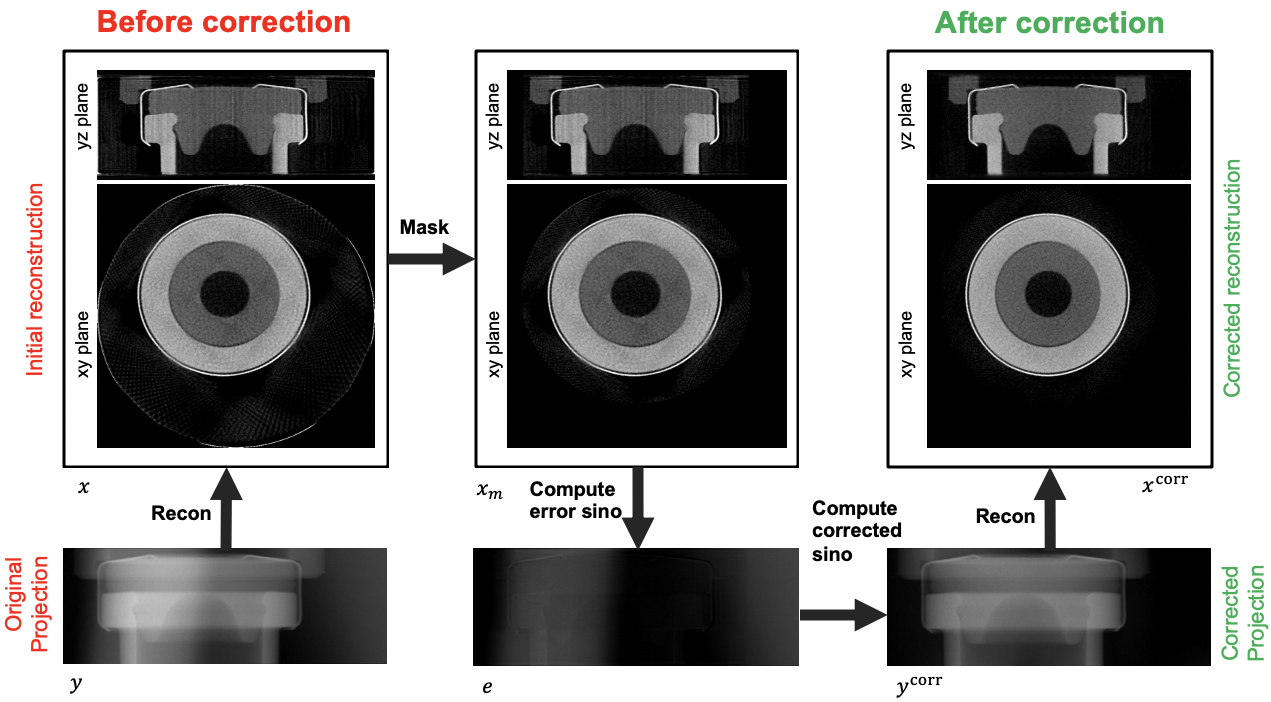}
\caption{Pipeline of the blind fixture correction in Algorithm~\ref{algo:blind_corr}.
The vertical stripes in the yz-plane of the reconstruction and the ring at the edge of the field of view in the xy-plane of the reconstruction have been rectified after performing the correction.
}
\label{fig:jigcorr}
\end{figure*}

Figure~\ref{fig:results_kwikpen_xy} shows a volume rendering of the reconstructed spring and a cross-section through it for four time-points and reconstruction methods FBP,  MBIR+4D-MRF, and multi-slice fusion.
The FBP reconstruction contains severe limited-angle artifacts.
MBIR+4D-MRF mitigates some limited-angle artifacts but some artifacts remain.
In contrast, multi-slice fusion mitigates most limited-angle artifacts.
The cross-sections of the spring in the multi-slice fusion reconstruction are more circular than the other methods, which align with our prior knowledge about the spring.
The fast compression of the spring causes the rings within the spring to move significantly within a time-point, resulting in the observed blur in the multi-slice fusion reconstruction.
Strong limited-angle artifacts in the other reconstructions mask this effect.

\section{Conclusion}

In this paper, we proposed a novel 4D X-ray CT reconstruction algorithm, multi-slice fusion, that combines multiple low-dimensional denoisers to form a 4D prior.
Our method allows the formation of an advanced 4D prior using state-of-the-art CNN denoisers without needing to train on 4D data.
Furthermore, it allows for multiple levels of parallelism, thus enabling reconstruction of large volumes in a reasonable time. 
Although we focused on 4D X-ray CT reconstruction for NDE applications, our method can be used for any reconstruction problem involving multiple dimensions.


\ifpeerreview
\else
\section*{Acknowledgment}
The authors would like to acknowledge support from Eli Lilly and Company under research project funding agreement
17099289.
Charles A. Bouman and Gregery T. Buzzard were supported in part by NSF grant CCF-1763896.
We also thank M. Cory Victor and Dr. Coralie Richard from Eli Lilly and Company for their assistance and guidance in setting up the residual seal force test experiment.
\fi

\appendices

\section{Correction for fixtures outside the field of view }
\label{sec:jigcorr}

\begin{algorithm}[!htb]
\DontPrintSemicolon
\SetKwInOut{Input}{Inputs}  
\Input{Original Sinogram: $y$,\\
System Matrix: $A$,\\
}
\KwOut{Corrected Sinogram: $ y^\text{corr} $}
$ x \gets \text{recon}(y, A) $\;
$ x^\text{m} \gets \text{mask}(x) $\;
$ e \gets y - A x^\text{m} $\;
$ p \gets \text{blur}(e) $\;
$ c \gets \argmin_{c \in \mathbb{R}} \| e - c p \|^2 $\;
$ y^\text{corr} \gets y - c p $\;
\caption{Blind fixture correction}\label{algo:blind_corr}
\end{algorithm}

Here we describe our correction for fixtures placed out of the field of view of the scanner.
As shown in Figure~\ref{fig:vial_setup}, the setup is held together by a fixture constructed of tubes and plates.
The tubes were placed outside the field of view of the CT scanner, thus causing artifacts in the reconstruction.
Our method performs a blind source separation of the projection of the object from that of the tubes.
Our blind separation relies on the fact that the projection of the tubes is spatially smooth.
This is true since the tubes themselves do not have sharp features and there is motion blur due to the large distance of the tubes from the rotation axis.

Algorithm~\ref{algo:blind_corr} shows our correction algorithm for the fixtures.
Figure~\ref{fig:jigcorr} illustrates the algorithm pictorially.
The initial reconstruction $x$ suffers from artifacts within the image and at the edge of the field of view.
We mask $x$ using a cylindrical mask slightly smaller than the field of view to obtain the masked image $x^m$.
This is done so that the majority of the artifacts at the edge of the field of view are masked but the object remains unchanged in $x^m$.
Consequently the error sinogram $ e = y - A x^m $ primarily contains the projection of the tubes with some residual projection of the object.
The blurring of $e$ filters out the residual object projection but preserves the spatially smooth projection of the tubes.
The corrected measurements $y^\text{corr}$ are found after performing a least squares fit.
The correction can be repeated in order to get an improved reconstruction $x$ and consequently an improved correction $y^\text{corr}$.

Figure~\ref{fig:jigcorr} shows the sinogram and reconstruction both before and after performing the blind correction.
Not only does the reconstruction after fixture correction remove the artifacts in the air region, but it also improves the image quality inside the object.
It can be seen that the vertical stripes in the object in the yz view of the reconstruction have been eliminated after performing the correction.

\bibliographystyle{IEEEtran}
\bibliography{references}

\end{document}